\documentclass[aps,nofootinbib,notitlepage]{revtex4}
\usepackage[english]{babel}
\usepackage{amstext,amsmath,amssymb,amsfonts,bbm}
\usepackage[latin1]{inputenc}
\usepackage{graphicx}
\usepackage{epstopdf}
\usepackage{amsthm}
\usepackage{tocvsec2}
\usepackage{enumerate}
\usepackage[T1]{fontenc}
\usepackage{color}
\usepackage{xcolor}
\usepackage{mathtools}
\usepackage{bbm}
\usepackage{braket}
\usepackage{tcolorbox}
\usepackage{natbib}
\usepackage{graphicx}

\usepackage{pgfplots}

\pgfplotsset{compat = newest}

\definecolor{darkgreen}{rgb}{0.0, 0.2, 0.13}
\definecolor{darkred}{rgb}{0.55, 0.0, 0.0}
\definecolor{carrotorange}{rgb}{0.93, 0.57, 0.13}
\usepackage{hyperref}
\hypersetup{
	colorlinks=true,
	linkcolor=blue,
	filecolor=magenta,
	urlcolor=cyan,
}
\usepackage[squaren,Gray]{SIunits}

\newcommand{\va}{\scriptscriptstyle}

\newcommand{\be}{\nopagebreak[3]\begin{equation}}
\newcommand{\ee}{\end{equation}}

\newcommand{\bee}{\nopagebreak[3]\begin{equation*}}
\newcommand{\eee}{\end{equation*}}

\newcommand{\ba}{\nopagebreak[3]\begin{eqnarray}}
\newcommand{\ea}{\end{eqnarray}}
\DeclareFontFamily{U}{rsfs}{}         % Formal Script            %
\DeclareFontShape{U}{rsfs}{m}{n}{<5> rsfs5 <6><7> rsfs7          %
  <8><9><10><10.95><12><14.4><17.28><20.74><24.88> rsfs10}{}     %
\DeclareMathAlphabet{\mathfs}{U}{rsfs}{m}{n}                     %
\newcommand{\mfs}[1]{\mathfs {#1}}                               %
\newcommand{\n}{{\nonumber}}

\newcommand{\sN}{{\mfs N}}

\newcommand{\sO}{{\mfs O}}

 \usepackage{braket}

\newcommand{\sbraket}[1]{\braket{\braket{#1}}}
\usepackage{etoolbox}

\makeatletter
\patchcmd{\frontmatter@abstract@produce}
  {\vskip200\p@\@plus1fil
   \penalty-200\relax
   \vskip-200\p@\@plus-1fil}
  {}
  {}
  {}
\makeatother
\begin{document}

\title{Planckian discreteness as seeds for cosmic structure}

\author{Lautaro Amadei}
%\email{sudarsky@nucleares.unam.mx}
\affiliation{{Aix Marseille Universit\'e, Universit\'e de Toulon, CNRS, CPT, Marseille, France}}

\author{Alejandro Perez}
\affiliation{{Aix Marseille Universit\'e, Universit\'e de Toulon, CNRS, CPT, Marseille, France}}
\date{\today}

\begin{abstract}
We propose a model of inflation driven by the relaxation of an initially Planckian cosmological constant due to diffusion. The model  can generate a (approximately) scale invariant spectrum of (adiabatic) primordial  perturbations with the correct amplitudes and red tilt without an inflaton. The inhomogeneities observable in the CMB arise from those associated to the fundamental Planckian granularity that are imprinted into the standard model Higgs scalar fluctuations during the inflationary phase. The process admits a semiclassical interpretation and avoids the trans-Planckian problem of standard inflationary scenarios based on the role of vacuum fluctuations. The deviations from scale invariance observed in the CMB are controlled by the self coupling constant of the Higgs scalar of the standard model of particle physics.  The thermal production of primordial black holes can produce the amount of cold dark matter required by observations.  
For natural initial conditions set at the Planck scale the amplitude and tilt of the power spectrum of perturbations observed at the CMB depend only on known parameters of the standard model such as the self coupling of the Higgs scalar and its mass.  
\end{abstract}
%-------------------------------------------------------------------------
%-------------------------------------------------------------------------

\maketitle

\definecolor{mycolor}{rgb}{0.122, 0.435, 0.698}

\section{Introduction}\label{intro}

Planck mass square, $m^2_p$, is the natural order of magnitude of the cosmological constant, yet its observed value is about $10^{-120}$ times that theoretical expectation. Such a huge discrepancy, referred to as the cosmological constant problem, is perhaps the most severe hierarchy problem of modern physics. The cosmological constant problem is often separated into two (possibly independent) questions: first why is the cosmological constant so small in Planck units, and second why does it have that special value. A related natural question that could connect the two is whether dark energy (the cosmological `constant') could actually change during the evolution of the universe. Namely, would it be possible to start with a large cosmological constant that dynamically evolves to its present value? 
If one writes the energy momentum tensor of regular matter and includes the dark energy component as ${\mathbf T}^{\va \rm TOTAL}_{ab}={\mathbf T}_{ab}+g_{ab} \Lambda/(8\pi G) $ then one has that Einstein's equation imply that
\be\label{inject}
\nabla_b \Lambda =- (8\pi G) \nabla^a{\mathbf T}_{ab}.
\ee
In other words, the only possibility (compatible with general relativity) of having $\Lambda$ change is the existence of diffusion of energy between the standard matter fields and dark energy. 
One could postulate such diffusion at a purely phenomenological level (for different proposals along these lines see \cite{Velten:2021xxw} and references therein). However, dark energy being associated with the gravitational properties of vacuum-spacetime, it is appealing to search for a more fundamental description that would presumably involve quantum gravity.

In this paper we will argue that the hypothesis of discreteness at the Planck scale provides a natural relaxation mechanism of the cosmological constant from its natural $m_p^2$ value. As a consequence of this, the universe undergoes a phase of exponential inflation during the initial part of the relaxation with a Hubble rate and scalar curvature of the order of the natural scales $m_p$ and $m_p^2$ respectively. The hypothesis of granularity at such scales naturally suggests a mechanism producing inhomogeneities in the matter sector that become eventually observable at the CMB. The process is different from the standard account where fluctuations in the CMB originate from vacuum fluctuations in the inflaton during inflation. The key difference is that in standard models of inflation one assumes that the relevant matter fields are in a suitable vacuum state (e.g. the Bunch-Davies vacuum in the De Sitter idealizations) as possible deviations from that special state have had time to dilute exponentially. Such vacuum state does not break the symmetries of the background spacetime geometry (assumed to be smooth to all scales in such accounts).  Indeed if we write an arbitrary matter field $\chi$ as $\chi=\chi_0+\delta\chi$ then $\braket{\psi| \delta\chi |\psi}=0$ (where $\chi_0$ is a constant background field that may or may not vanish). In our account, instead, the background geometry is not homogeneous and isotropic at short scales due to Planckian granularity expected from quantum gravity. Such Planckian discreteness interacts with matter fields and produces inhomogeneities such that $\braket{\psi| \delta\chi |\psi}\not=0$ from their birth at horizon crossing. In our account we assume that we can treat these fluctuations semiclassically from the time of horizon crossing on. The details of the physics at shorter scales is expected to be described by a suitable quantum gravity theory; we only assume that the result of that physics is the production of inhomogeneities at horizon crossing in a way that can be approximated by a stochastic process (the details of this will be presented in Section \ref{sf}). In the standard inflationary accounts one needs to assume that the analog of the Bunch-Davies vacuum holds true to scales much shorter than the Planck scale (this is usually called the trans-Planckian problem).

One can also note that the expectation value of the energy momentum tensor of the matter fields $\braket{\psi|T_{ab}|\psi}$ in the suitable quantum state $\ket{\psi}$ breaks homogeneity and isotropy in our case (because Planckian granularity leaves imprints in the primordial state of matter $\ket{\psi}$) while  $\braket{\psi|T_{ab}|\psi}$ is perfectly homogeneous and isotropic in standard accounts of inflation.  This feature of the standard approach raises a key question leading to tensions in the context of the interpretation of quantum mechanics in a closed universe: {\em how or what produces the symmetry breaking (of the FLRW symmetries) from the primordial stage to the final inhomogeneous state of the CMB?}
Our perspective eliminates the question as the FLRW symmetries are broken by the Planck scale granularity from the onset.  The reader will find that a certain number of assumptions are necessary to reach quantitative predictions in our model (for an attempt of discussing some of the open issues that remain see Appendix \ref{openissues}). The validity of these assumptions is of course central for the quantitative validity of the predictions. However, independently of this, we believe that our work is also valuable as a proof of concept showing the existence of an alternative view on the origin of structure in the universe. 
 
 \subsection{Apologia of unimodular gravity}
 
We would like to come back to the equation \eqref{inject}  and point out that it arises naturally in the context of unimodular gravity in a way that, we believe, has some additional conceptual value in view of the previous discussion. Thus let us explore it in some detail as the perspective  it suggests motivates a feature of the model that we introduce below.  
The action of unimodular gravity is 
\be\label{action}
S=\int\left(\sqrt{g} \ {\mathbf R} +\lambda \left[\sqrt{g}-v^{(4)}\right]\right) dx^4+S_{m},
\ee
where $S_m$ denotes the action of matter fields, and 
\be\label{4v}{\mathbf v}^{(4)}\equiv v^{(4)} dx^0\wedge dx^1\wedge dx^2\wedge dx^3\ee
is a background four-volume form, and $\lambda$ is a Lagrange multiplier imposing 
that the metric volume density equals the background one. The presence of the four-volume background structure breaks, in unimodular gravity, the diffeomorphism symmetry of general relativity down to volume-preserving diffeomorphisms, whose generators are represented by the vector fields $\xi^a$ with vanishing expansion $\theta$, namely 
\be\label{ff}
\theta\equiv \nabla_a\xi^a=0~.
\ee
Such infinitesimal generators of volume preserving diffeomorphism are completely characterized by arbitrary 2-foms $\omega_{ab}$ via the relation $\xi^a=\epsilon^{abdc}\nabla_b\omega_{cd}$. 

Invariance of the matter action under the so restricted diffeomorphisms relaxes the usual constraints on the divergence of the energy momentum tensor. Recall that full diffeomorphism invariance of the matter action implies energy momentum conservation (see for instance \cite{Wald:1984rg}). Therefore, in order to find the 
new constraints on energy conservation one must set to zero the variation of the matter action under volume preserving diffeomorphisms under the assumption that the matter field equations hold. Namely, the new condition reads
\ba
0&=&\delta S_m=\int_M \sqrt{-g} {\mathbf T}_{ab} \nabla^a\xi^b dx^4\n \\ &=&-\int_M \sqrt{-g} \nabla^a{\mathbf T}_{ab} \xi^b dx^4=\int_M \sqrt{-g} \nabla_c(\nabla^a{\mathbf T}_{ab}  \epsilon^{bcde})\omega_{de} dx^4,
\ea 
where we integrated by parts twice and we have assumed that fields vanish at infinity. If we define \be {\mathbf J}_b\equiv (8\pi G )\nabla^a{\mathbf T}_{ab},\label{enema}\ee the previous condition, which should be valid for arbitrary $\omega_{ab}$,  implies  
\be\label{integ}
d{\mathbf J}=0,
\ee
or locally
\be\label{toma}
{\mathbf J}_a=\nabla_a Q,
\ee 
where $Q$ is some scalar.
Therefore, the background volume structure---that partially breaks diffeomorphims (down to volume preserving ones)---allows for violations of energy momentum conservation demanding only that the energy-momentum violation current ${\mathbf J}_b$ be closed.  The gravitational field equations that follow from the previous action are simply the trace-free part of Einstein's equations, namely  
\begin{equation}\label{TraceFreeEinsteinEquation}
	{\mathbf R}_{ab} - \frac{1}{4} {\mathbf R} g_{ab} = {8 \pi G} \left( {\mathbf T}_{ab} - \frac{1}{4} {\mathbf T}g_{ab} \right),
\end{equation}
which, using the integrability condition (\ref{integ}) and the Bianchi identities, can be rewritten as \cite{Josset:2016vrq}
\begin{equation}\label{TraceFreeEinsteinEquation2}
{\mathbf R}_{ab} - \frac{1}{2} {\mathbf R} g_{ab} +\underbrace{\left[\Lambda_{ 0} + \int_{\ell} {\mathbf J}\right] }_{\Lambda}g_{ab}= 
{8 \pi G}  {\mathbf T}_{ab} ,
\end{equation}
where $\Lambda_0$ is a constant of integration, and $\ell$ is a one-dimensional path from some reference event to the spacetime point where the equation is evaluated. Thus, if not vanishing, the energy-violation current $\mathbf J$ is the source of a term in Einstein's equations satisfying the dark energy equation of state; while if $\mathbf J=0$ we simply recover the field equations of general relativity with a cosmological constant $\Lambda_0$ (a property aready pointed out by Einstein \cite{Einstein1919Spielen-Gravita} as indicating the possibly non-fundamental nature of the cosmological constant). In relation to this, another very appealing feature of unimodular gravity is that quantum field theoretic vacuum energy does not gravitate \cite{Weinberg:1988cp, Ellis:2010uc}, for vacuum fluctuations only contribute to the trace part of ${\mathbf T}_{ab}$ not entering into the field equations \eqref{TraceFreeEinsteinEquation}. Finally, and highly remarkably, aside from new physics in the dark matter sector, unimodular gravity is completely equivalent to general relativity \cite{Ellis:2013eqs}, and passes all the known tests of Einstein's theory. Unimodular gravity is, therefore, a very conservative modification of general relativity.

What is the role of the background four volume structure? Why should one accept such weakening of the principle of general covariance (breaking diffeomorphism down to volume-preserving diffeomorphisms)? We are guided on this issue by the perspective that the smooth classical field description of general relativity and quantum field theory is an approximation  (an effective description) of a fundamental physics expected to be discrete at the Planck scale. Compatibility with Lorentz symmetry suggests that such discreteness would have to be realized by the existence of some sort of four-volume elementary building blocks. These basic spacetime elements would naturally produce a background 4-volume structure in the long wavelength effective description and justify the use of unimodular gravity for low energies. Such background structure would break diffeos down to volume preserving diffeos in an effective description where the volume elements are not dynamically included \cite{Anderson:1971pn}.  At the more fundamental level (i.e., in terms of the quantum gravity physics describing the dynamics of such elementary notions) no background structures should be preferred, a priori, and full covariance would be reestablished. 

 Indications of this physical hypothesis come from different indirect sources that we now mention. 

First, let us come back to the discussion of the symmetries of unimodular gravity and recall that under general diffemorphisms the metric changes as $\delta g_{ab}=2\nabla_{(a} \xi_{b)}$ where $\nabla_{(a} \xi_{b)}=\frac{\theta}{4}  g_{ab}+\sigma_{ab}$
when decomposed in its trace and trace-free parts.
Unimodular gravity remains invariant under the smaller group of volume-preserving diffeomorphisms which are characterized infinitesimally 
by vector fields $\xi^a$ for which $\theta=0$ (equation \eqref{ff}). 
Thus, the broken diffeomorphisms in unimodular gravity are those that send the metric $g_{ab} \to (1+\frac{\theta}{4}) g_{ab}$ which coincide with infinitesimal conformal transformations $g_{ab}\to \Omega^2 g_{ab}$ as far as the metric is concerned. Therefore, when the field equations hold, 
conformal transformations and the broken symmetries of unimodular gravity are the same in the matter sector. Thus, one would expect unimodular gravity to emerge as the natural  effective description of gravity in situations where scale invariance is broken by the microscopic discreteness scale associated to quantum gravity scale and those of the fundamental probing matter \footnote{There is a remarkable paper by Anderson and Finkelstein \cite{Anderson:1971pn} where a very similar conceptual path leads to unimodular gravity from the assumption of the existence of a fundamental scale breaking conformal invariance. Although in their analysis they do not discuss the possibility of diffusion that is one of the key features of our approach.  We were not aware of this paper and thank T. Jacobson for pointing it out to us.}.

This is precisely what the structure of quantum field theory on curved spacetimes suggests in the way the UV (potentially divergent) contributions to the renormalization of the energy-momentum tensor break scale invariance: consideration of the ambiguities associated with the definition of the expectation value of the energy momentum tensor in quantum field theory and their (anomalous) breaking of scale invariance \cite{ Wald:1995yp} can be argued to indicate the preferred role of unimodular gravity in semiclassical considerations. 

The previous discussion based on pure symmetry considerations can be made very concrete. Still in the context of the renormalization of the (expectation value of the) energy momentum tensor in quantum field theory on curved spacetimes, the existence of a well defined regularization can be shown via the Hadamard subtraction prescription where---for the simple case of a Klein-Gordon field $\phi(x)$---one defines $\braket{T_{ab}}$
by considering the coincidence limit $x\to y$ of a suitable expression depending on the two-point distribution
%\be
%\braket{T_{ab}(x)}=\lim\limits_{y\to x}\left[\nabla^x_{(a}\nabla^y_{b)} F(x,y) -\frac{1}{2} g_{ab}(\nabla^x_c\nabla^c_y +m^2) F(x,y)\right]
%\ee
\be
F(x,y)=\braket{\phi(x)\phi(y)}-H(x,y),
\ee 
where $H(x,y)$ is a Hadamard bi-distribution constructed such that $F(x,y)$ is smooth in the coincidence limit, and such that 
it satisfies the field equations in its first argument. Obstructions to get $H(x,y)$ to satisfy the field equations in the second argument
imply---when replacing in the suitably defined point split regularization of the energy momentum tensor---that in the coincidence limit $x\to y$ \be\label{yep}
\nabla^a\braket{T_{ab}(x)}=\nabla_b Q,
\ee 
for $Q$ dependent on the local background geometry curvature but not on the state of the quantum field. Thus, the simple regularization of the UV divergences leads to the violation of energy momentum conservation of the 
form compatible with the symmetries of unimodular gravity \eqref{toma}.

In order to make the formalism compatible with the usual semiclassical equations one performs an additional step `by hand' \cite{ Wald:1995yp}
and defines the {\em renormalized} expectation value $\braket{T_{ab}(x)}_{\rm ren}$ by  \be\label{prbrb}
\braket{T_{ab}(x)}_{\rm ren}\equiv \braket{T_{ab}(x)}-Q\,  g_{ab},
\ee  which, in the case of conformal quantum fields,  introduces an anomalous trace (see for instance \cite{Birrell:1982ix, Wald:1995yp}, and \cite{Fabbri:2005mw} for a very detailed and transparent presentation which is available in 2-dimensions). Our previous discussion shows that this anomaly is more naturally interpreted as a violation of energy-momentum conservation \eqref{enema} satisfying the unimodular restriction \eqref{integ}.  

Interestingly,  the semiclassical gravity dynamics defined using the trace free Einsteins equations of unimodular gravity with sources given by  $\braket{T_{ab}}$ (violating conservation as in \eqref{yep}), and the one implied by the standard Einstein's equations with sources defined by \eqref{prbrb} coincide. In this sense, the trace anomaly is equivalent to a diffeomorphism anomaly where diffeomorphisms are broken---by QFT vacuum fluctuations---down to volume preserving diffeomorphism. Quantum fields and their fluctuations around a preferred (`vacuum') state are sensitive, in this sense, to an underlying four-volume structure.
Finally it is worth pointing out that---even though there are ambiguities in the definition of $\braket{T_{ab}}$ encoded in the possibility of adding a locally conserved
tensor $t_{ab}$ constructed from the metric variation of a Lagrangian constructed out of $R^2$ and $R_{ab} R^{ab}$ \cite{Wald:1995yp}---the relevant `diffusion' term $\nabla_b Q$ is, to our knowledge, unambiguously  defined in the present context. All this strengthen the view that unimodular gravity---with the non trivial diffusion \eqref{toma} effects that it offers---is a natural effective description emergent from the underlying UV structure of spacetime and matter expected to be described by quantum gravity \footnote{The possibility of a relaxation mechanism of a positive cosmological constant via the back reaction of infrared graviton modes (IR effects)  was put forward by Tsamis and Woodard in \cite{TSAMIS1993351} and further explored in the case of scalar contributions by Brandenberger \cite{Brandenberger:2002sk}.}.
  
Discreteness at the Planck scale (or, more precisely, the existence of microscopic degrees of freedom not accounted for in an effective field theory description) is suggested also by the physics of black holes in the semiclassical regime \cite{Perez:2017cmj}. Black holes behave like thermodynamical systems in quasi-thermal equilibrium with an entropy given by
\be
S_{BH}=\frac{A}{4\ell_p^2},
\ee
where $A$ is the corresponding black hole horizon area. This formula suggests the existence of microscopic degrees of freedom at the Planck scale, $\ell_p$, responsible for such huge entropy. Arguments that take these microstates as fundamental and derive from them an effective description of gravity (as an equation of state \cite{Jacobson:1995ab}) lead---not to Einstein's equations as it is often improperly stated, but rather---to the trace free Einstein's equation (\ref{TraceFreeEinsteinEquation}) of unimodular gravity. 

As realized by Hawking in the 70's, black holes evaporate via the emission of thermal radiation and thus seem to destroy the information in the initial pure state that lead to their formation in violation of the expected unitarity of quantum gravity. This is the famous information paradox. It is important to point out here that the presence of microscopic degrees of freedom at the Planck scale offer a natural resolution of the paradox. Indeed, if these hidden degrees of freedom can interact with the low energy ones appearing in our effective field theory formulations then quantum correlations with the microscopic Planckian structure can be established via such interactions. This is particularly relevant in the context of black hole formation and evaporation where low energy excitations falling into the black hole are forced by the gravitational field to interact with the Planck scale a finite proper time after horizon crossing as they approach the classical singularity (as implied by the singularity theorems). This offers a natural channel for purification of the Hawking radiation \cite{Perez:2014xca} in a way that finds simple analogies in everyday systems where information is degraded (or made unavailable) because unitary evolution leads to decoherence with microscopic molecular type of degrees of freedom (e.g. when a newspaper information is lost into molecular chaos after burning the paper). This perspective can be explicitly tested in toy models in quantum cosmology illustrating the mechanism see \cite{Amadei:2019ssp, Amadei:2019wjp}). For a recent discussion of this view and a related one see \cite{Perez:2022jlm}.
The point we stress here is that such theoretical considerations are not disconnected from the present discussion in cosmology as they give extra strength to the hypothesis of Planckian discreteness which will play, in our model,  a central role in the genesis of inhomogeneities observed in the CMB.  

Unimodular gravity also arises naturally from quantum gravity approaches where spacetime is emergent from 4-dimensional discrete building blocks (which are responsible for the existence of a preferred background four-volume \eqref{4v}). A concrete example of this is the role of unimodular gravity as the effective description of gravity in the causal set approach \cite{Bombelli:1987aa}. Noisy interaction with four volume events appears as the natural relativistic generalization of spontaneous localization models \cite{Bedingham:2010hz} that modify quantum mechanics by introducing dynamical collapse \cite{Ghirardi:1989cn,Ghirardi:1985mt}. This perspective was relevant in motivating the use of unimodular gravity in \cite{Josset:2016vrq} where observational bounds on the free parameters of some of such models where constrained by cosmological observations.  It is possible that these, apparently independent directions, could be connected at a more fundamental level. We will not 
pursue this idea here, for further reading and applications to cosmology see \cite{Canate:2013isa} and references therein.

The existence of microscopic degrees of freedom which are not captured in our smooth field theoretic approximations conveys the idea that diffusive effects could be present which, in unimodular gravity, can be accounted for phenomenologically in terms of a non vanishing current ${\mathbf J}_b$ (as long as (\ref{integ}) are satisfied). This perspective, which is the one we follow in this work, was already taken in \cite{Perez:2017krv, Perez:2018wlo} where (with the assumption that the initial cosmological constant $\Lambda_0$ in \eqref{TraceFreeEinsteinEquation2} is vanishing in the early universe) a cosmological constant emerges from the noisy diffusion of energy from the low energy matter sector into the Planckian regime during the electroweak transition. Remarkably,  the model reproduces the observed value of the cosmological constant today without fine tuning \footnote{The model links the two mysteriously  small scales in fundamental physics---the EW scale $m_{\rm ew}$ and the cosmological constant---with the gravity scale $m_p$: the small number $(m_{\rm ew}/m_p)^7\approx 10^{-120}$ emerges from the calculation as a result of the diffusive physics involved \cite{Perez:2017krv, Perez:2018wlo}. The results of the present paper reinforces the relationship between dark energy physics and electroweak physics due to the key role that the Higgs scalar will play in what follows.}.

\subsection{{\em Background implications:} Relaxation of the cosmological constant}

Building on this, here we explore the possibility that the perspective offered by unimodular gravity (as an effective description emerging from fundamental discreteness) could help addressing the first part of the cosmological constant problem. We would like to investigate the cosmological implications of having an initial cosmological constant that starts with its natural Planckian value $\Lambda_0\approx m^2_p$, and then relaxes to zero via diffusion into the matter sector mediated by the hypothetical granular structure at the Planck scale associated with the emergence of the preferred four volume structure of unimodular gravity at low energies.  

The analogy with usual dissipative systems suggests a natural model where $\Lambda$ relaxes exponentially in time. Even if time is an elusive notion in general relativity,  when it comes to applying the theory to cosmology the situation is drastically different in unimodular gravity (for a more general discussion see \cite{Unruh:1988in, Smolin:2010iq}). This is so thanks to the existence of the preferred 4-volume structure that singles out a preferred (up to rescaling by a constant) notion of time: $4$-volume time. Such a time variable can be put in direct correspondence with a dimensionless notion associated with the counting of elementary Planckian volume elements `created' during the cosmological evolution. All this  provides a natural time notion emerging from the hypothesis of discreteness in terms of which the relaxation of $\Lambda$ will be defined.  

To make the previous statement precise we now focus our attention to (spatially flat) Friedmann-Lema\^\i tre-Robinson-Walker (FLRW) cosmology (homogeneous and isotropic cosmology). The assumption of spatial flatness simplifies the discussion that follows yet it is probably not essential. Thus the spacetime metric       
is given by 
\be\label{metric}
ds^2=-d\tau^2+a^2(\tau) d\vec x^2,
\ee
where $\tau$ is the proper time of co-moving observers. The rationale dictating that the diffusion is sourced by the $4$-volumetric granularity of spacetime suggests the natural time for the diffusion process (and associated relaxation of $\Lambda$) to be proportional to the number of spacetime grains encountered or `created' during the evolution of the universe identified with the  elapsed four volume. More precisely, consider an initial fiducial cell of co-moving coordinate volume $\ell_p^3$ expanding while the universe expands. The four volume of its world tube---divided by a reference volume scale $\ell_{\rm U}^3$ in order to get time units---is given by
\be\label{titita}
t_p=\frac{\ell_p^3}{\ell_{\rm U}^3}\int a^3 d\tau.
\ee
In order to fix the scaling degeneracy of four volume related time notions we take $\ell_{\rm U}=\ell_p$, which produces the so-called unimodular time variable $t$ defined as 
\be\label{12}
dt=a^3 d\tau, 
\ee 
which  turns the metric \eqref{metric} into  $ds^2=-a^{-6} dt^2+a^2 d\vec x^2$. This time choice is imposed to us in unimodular gravity by the constraint $\det{|g|}=1$ derived from the variations of the action \eqref{action} with respect to the Lagrange multiplier $\lambda$ in natural coordinates where $v^{(4)}=1$. 

The question we explore in this paper is what is the natural phenomenology that follows from the assumption that $\Lambda$ decays exponentially in this (number of Planck four volume elements) time, thus we postulate that
\be\label{lala}
\Lambda(t)=\Lambda_0 \exp(-\beta m_p t),
\ee
with  $\beta$ a dimensionless constant  and $\Lambda_0\sim m_p^2$.   Note that $n_p\equiv m_p t$ in the previous expression can be interpreted as the number of Planckian $4$-volume elements `created' during the cosmological expansion
out of the primordial initial cell. Note also that the time variable as defined in \eqref{titita} is not unique as it can be modified by rescaling $\ell_p \to \lambda \ell_p$. The phenomenology of this paper remains the same if simultaneously we rescale $\beta\to \beta/\lambda^3$ in \eqref{lala}.  This freedom can be encoded in the choice of $\ell_{\rm U}$ in \eqref{titita}.  We will see that the parameter $\beta$ will basically control the number of e-folds of inflation before reheating. The only requirement we will find, when comparing predictions of the model with observations, is that $\beta$ has to be sufficiently small. However, its precise value does not  affect the type of observable features we explore in the model. A possibility of identifying a fundamental mechanism fixing this freedom and, simultaneously, rendering the value of $\beta$ more natural will be discussed in Section \ref{openissues}. The ansatz \eqref{lala} is certainly speculative at the present stage of understanding of quantum gravity; however, we will see that it leads to an alternative view on the origin of structure in the late universe. This can be taken as a proof of concept of a different perspective whose assumptions may be weaken  with future investigations.

We will see that \eqref{lala} implies, due to the non trivial relation between the (four volume) time $t$ and co-moving time $\tau$ encoded in equation \eqref{12}, that  the universe undergoes a phase of exponential expansion in cosmic time $\tau$  lasting as long as $\beta m_pt<1$ (a quasi De Sitter inflationary phase). This is explicitly seen in the dependence of $\Lambda$ with $a$ derived in \eqref{endy}.   For sufficiently small $\beta$ this inflationary phase can be long enough to resolve both the horizon and the flatness problems independently of the initial conditions\footnote{The independence of initial conditions should be taken with the same grain of salt as when one reads similar statements in the inflationary literature. More precisely, one can only make a statement of this sort once one assumes that the FLRW approximation is a good one to describe the observable universe. This  is clearly a severe restriction of the phase space of general relativity as it is often emphazized by Penrose \cite{Penrose:1994de}, and of course a very important problem that we will leave aside of the present discussion. } for matter fields and the energy injection encoded in equation \eqref{inject}. We will discuss this in more detail in Section \ref{bg}. 

\subsection{{\em Perturbation implications:} {Inhomogeneities sourced by Planckian granularity}}

The conceptual framework of unimodular gravity naturally suggests the possibility for a form of diffusion between the matter degrees of freedom and the dark energy sector (representing an evolving cosmological constant ). The rational behing all this is the existence of hidden Planckian degrees of freedom which, in the effective low energy description of unimodular gravity,  are capable of storing energy in the form of dark energy to be eventually released into the degrees of freedom of matter (a mechanism driven by quantum gravity and here assumed as a phenomenological hypothesis to lead to the relaxation of $\Lambda$ as in \eqref{lala}). However, as this dark energy is freed by Plankian grains of spacetime, we can envisage the possibility that inhomogeneities would arise in the matter sector at around the fundamental scale which, during the inflationary period, is close (as shown in Section \ref{bg}) to the Hubble rate. At present one cannot describe this process from fundamental principles. Thus we will represent it by a Brownian type of process, i.e a stochastic process generating small perturbations of certain background fields with a probability distribution satisfying the only requirement of homogeneity.

This view offers an interesting possibility for a mechanism of structure formation where the nearly scale invariant scalar density fluctuations observed in the CMB will be shown to arise from the steady injection of energy from the Brownian-like diffusion of fundamental Planckian granularity into the perturbations at the Hubble scale during the De Sitter phase. We will see that the semiclassical description of such diffusion leads to stochastic inhomogeneities compatible with cosmological observations.  Scale invariance follows from the self similarity of the diffusion process that is granted by the exponential expansion of the background during the De Sitter phase (due to the slow relaxation of $\Lambda$ as in \eqref{lala}). 

Thus the mechanism producing inhomogeneities presented here is fundamentally different from the standard account that associates inhomogeneities to quantum fluctuations of the inflaton. Here we propose an active mechanism where the fundamental quantum granularity induces semiclassical inhomogeneities in the mean field value of the Higgs scalar. Why the Higgs scalar instead of any other field in the standard model of particle physics (which we assume to be valid up to close to the Planck scale)?  To answer this question first note that inhomogeneities are expected to be intrinsically present at the Planck scale according to several approaches to the fundamental theory. However, compatibility with Lorentz invariance implies that such hypothetical granularity cannot be seen as an underlying lattice-like structure selecting a preferred frame \cite{PhysRevLett.93.191301}. Instead, discreteness at the Planck scale must have physical manifestations when  suitable massive (hence scale invariance breaking) degrees of freedom interact with the quantum geometry (massless fields cannot be sensitive to granularity as their light-like excitations cannot define a frame, their own rest frame, with respect to which the notion of Planck scale would be meaningful). At high enough energies, the only scale invariant breaking degree of freedom in the standard model of particle physics is the Higgs scalar and this is  the reason why the Higgs is the right degree of freedom that can carry the imprints of granularity.   A natural order parameter of the magnitude of the strength of this effect is naturally given by 
\be\label{conficinfi}
\gamma_{\rm H}=\frac{m_{\rm H}}{m_p}\approx 10^{-17},
\ee  
where $m_{\rm H}$ is the Higgs mass.
  
Thus in our model the fundamental inhomogeneities leave their imprint on the expectation value of the Higgs scalar (assumed to have, as the cosmological constant, Planckian initial value): as a consequence the Higgs scalar is not in a homogeneous and isotropic vacuum state but rather in an inhomogeneous  excited semiclassical state.   The De Sitter exponential expansion during the inflationary phase dilutes standard forms of matter; however, this is not the case for the zero mode of a scalar field and the inhomogeneities produced on it (long wavelength modes in the scalar field are frozen by the rapid expansion). During the inflationary phase the UV Planckian inhomogeneities are expanded to the large cosmological scales where they become the seeds for the formation of structure observable in the power spectrum of perturbations on the CMB. There is no symmetry breaking of the FLRW symmetries, no need for quantum to classical transition,  inhomogeneities are present from the beginning in the microscopic quantum gravitational structure of spacetime and matter. The decaying cosmological constant and its inflationary effect, bring these up to our scales.

In our view, at the conceptual level, the new perspective is an improvement of the standard picture in two ways: On the one hand it offers a possible resolution of the so-called trans-Planckian problem because no assumptions about the validity of standard quantum field theory as well as linearized gravity are necessary at length scales below the Planck scale. However, we still have to assume that standard semiclassical tools are accurate for length-scales slightly longer than the Planck scale where our perturbations are born (this is suggested by studies of simplified quantum cosmology models, see \cite{Ashtekar:2011ni}). On the other hand our approach eliminates the conceptual difficulties \cite{Perez:2005gh} associated  with thinking of the perturbations as originating in vacuum fluctuations of the inflaton in relation to the measurement problem in quantum mechanics and applications of its Copenhagen interpretation applied to the universe as a whole.
Of course our perspective does not eliminate all conceptual problems as the mechanism we invoke is deeply rooted in a quantum gravity rationale: it urges one to try to understands better the deep Planckian regime.

Finally we study the possibility that primordial black holes could be created thermally at the end of the inflationary era during the reheating phase that in our model raises the temperature to close to the Planck temperature. A key assumption here is that there are stable primordial black holes with masses close to the Planck mass. Note that even when this is suggested by general quantum gravity considerations in various contexts, it is a very natural possibility in a quantum gravity theory where the Planck energy is the fundamental scale. We show that natural estimates based on dimensional analysis lead to the correct order of magnitude densities necessary  to account for dark matter today without fine tuning.  We explain this in detail in Section \ref{de}.

The paper is organized as follows. In the  Section \ref{bg} we describe the 
dynamics of the background geometry driven by the relaxing cosmological constant  \eqref{lala}. In Section \ref{sf} we present the proposed mechanism for the generation of nearly scale invariant scalar density fluctuations. We confront the predictions of the minimalistic model (that assumes the validity of the standard model of particle physics all the way to the Planck scale) with the relevant observational data coming from the CMB. In section \ref{de} we analyze the possibility that primordial black holes (generated during the diffusion process or via thermal fluctuations at reheating) could account for the dark energy content of the universe.  We conclude the paper with a discussion Section \ref{dis}. Appendix \ref{WT} contains a proof of the so-called Weinberg theorem showing the existence of adiabatic solutions of the perturbation equations. This theorem is key in understanding the link between perturbations generated during inflation and the CMB observations.  In our context the theorem is a handy shortcut specially adapted to the dynamical description of the relevant consequences our stochastic process for the generation of inhomogeneities equivalent of the (more generally used) Mukhanov-Sasaki formalism in the description of standard inflationary theory of perturbations. We believe that our proof of the Weinberg theorem (even when the same in spirit as the one found in \cite{Weinberg:2003sw}  or in his well known textbook \cite{Weinberg:2008zzc}) is more direct and could be helpful for interested readers. In Appendix \ref{BD} we compare our mechanism for the generation of inhomogeneities with the standard paradigm.  Some of the various issues opened by our perspective are considered in Appendix \ref{openissues}.

\section{Background dynamics}\label{bg}

In this section we study the dynamics of the homogeneous and isotropic FLRW geometry  \eqref{metric} and homogeneous and isotropic matter components
evolving on it. The primordial cosmological constant (or dark energy component) relaxes according to \eqref{lala} and, we assume, that the energy released feeds (as implied by equation \eqref{inject}) a radiation component---represented by a homogeneous and isotropic perfect fluid with equation of state $\rho=3P$---whose initial value is $\rho_0$. This radiation fluid represents the massless degrees of freedom in the matter sector which (according to the standard model at high values of the Higgs scalar $vev$) are basically only photons and gravitons.
Equation \eqref{inject}  will take the form of a continuity equation with non trivial interactions between the radiation and dark energy fluid components.  Naturalness of initial conditions at the Planckian regime suggests $\rho_0\sim m_p^4$.
In addition we have the Higgs scalar field that is assumed to start off in a semiclassical state with expectation value $\phi_0(0)\sim m_p$.
However, the Higgs in such high energy initial state decays into particles of the standard model producing further interaction terms in the continuity equation (now between the Higgs energy-momentum tensor and the radiation). We will see that these interactions are weak in the regime of interest and that a semiclassical description is available.
Thus, in spite of the apparent complexity of the situation one can actually use analytic methods to get a quantitative picture of the relevant features of the dynamics of the background fields which fits well the numerical simulations (whose results we report in Figure \ref{fig:loglog}). 

We show in this section that, initially, the dynamics is dominated by the decaying cosmological constant---in a way that is independent of the other matter components and their initial conditions---producing an inflationary era of the De Sitter type that can last a large number of e-folds as in standard inflationary models \cite{Martin:2013tda}.  The e-folds of inflation are controlled by the parameter $\beta$, CMB observations require this number to be larger than a minimum value but they do not constrain it otherwise. Thus the free parameter $\beta$ is degenerate in this sense.         

\subsection{Quasi De Sitter phase from the relaxing $\Lambda$}\label{desi}

\begin{figure}[h]  
\centerline{\hspace{0.5cm} \(
  \begin{array}{c}
\includegraphics[height=8cm]{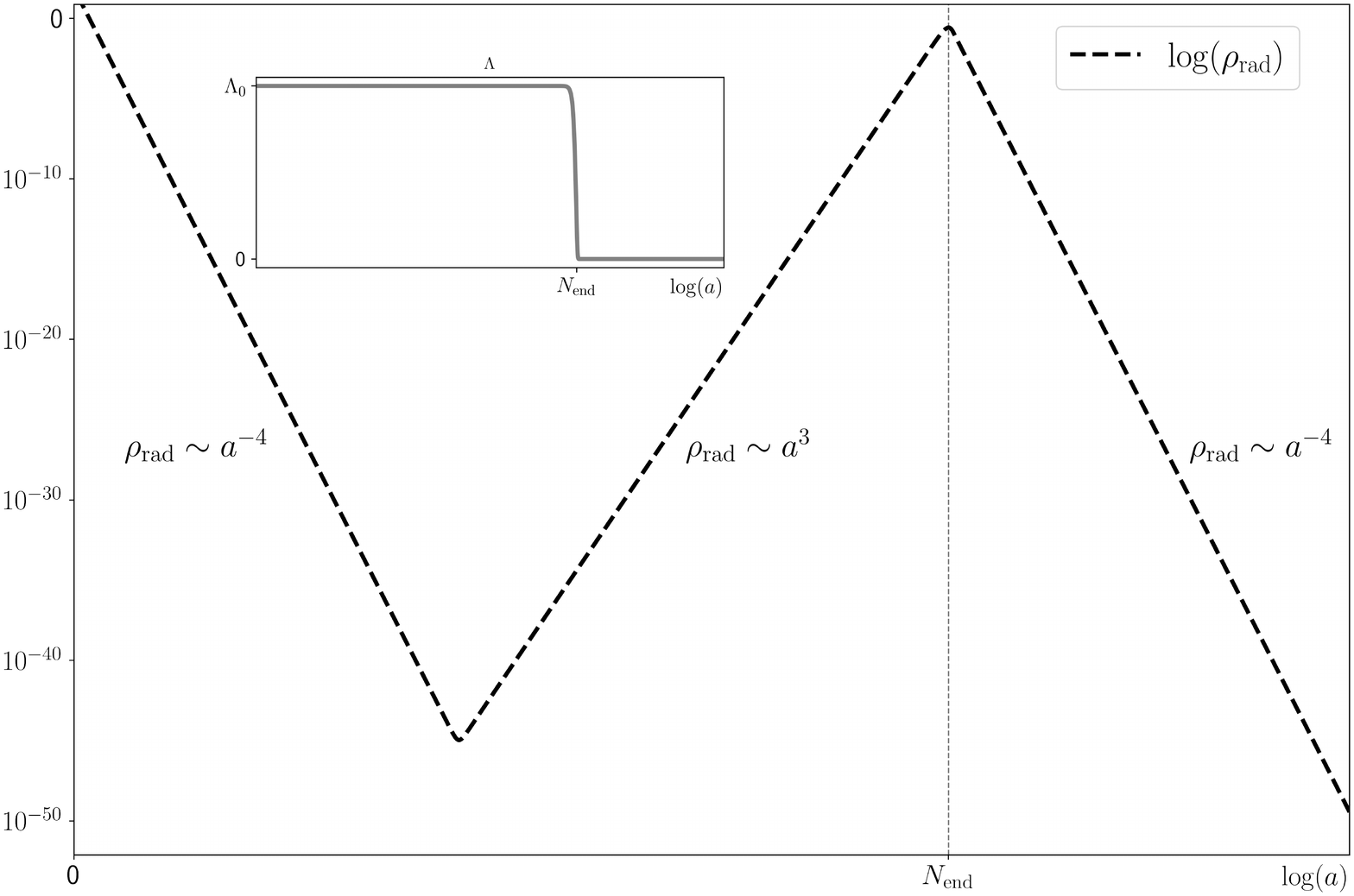} 
\end{array} \)} 
\caption{Numerical solution of \eqref{eq2} with $\beta = 10^{-80}$, we plot the cosmological constant $\Lambda$ (inserted panel in linear scale), and the radiation energy density $\rho_\rad$ (in log scale) in terms of the number of e-folds $\log(a)$. $\Lambda$ behaves effectively as a constant until about when condition \eqref{condimentos} is satisfied and abruptly decays to zero thereafter. The radiation density decays exponentially from its initial Planckian value until the energy injection from the relaxation of $\Lambda$ starts winning over the expansion. By the end of inflation radiation density grows back ({\em reheating}) to about Planckian density again (the reheating temperature is estimated in equation \eqref{fine}). } \label{fig:loglog}  
\end{figure}

We assume that the matter content of the universe is well represented by a perfect fluid,
\be
T_{ab}=\rho u_a u_b+P (g_{ab}+u_au_b),
\ee 
where $u_a$ is the $4$-velocity of co-moving observers, and $\rho$ and $P$ are the energy density and pressure in the  co-moving frame.
In terms of 4-volume (unimodular) time $t$ (see equation \eqref{12}) the Friemann equation becomes
 \begin{equation}\label{eq1}
    \begin{aligned}
   a^{4} {(a^\prime)^2} &= \frac{8 \pi G}{3}\rho + \frac{\Lambda(t)}{3}  ,
%   \\
%    a^{2} \frac{d}{ds} \left( a^{3} a^\prime \right) = - \frac{8 \pi G}{6} (\rho + 3 P) + \frac{\Lambda(t)}{3} \\
%    \dot{\rho}& + 3 \frac{a^\prime}{a} (\rho + P)  = - \frac{\dot{\Lambda}(t)}{8 \pi G}
    \end{aligned}
\end{equation}
where $\prime$ denotes derivatives with respect to unimodular time $t$, and  $\Lambda(t) =  \Lambda_{0} e^{-\beta m_p t}$ is the decaying cosmological constant depending on the  free parameter $\beta$, and from now on we normalize the scale factor so that $a(0)=1$. 
The Raychaudhuri equation is
\be\label{222} a^{2} \frac{d}{dt} \left( a^{3} a^\prime \right) = - \frac{4 \pi G}{3} (\rho + 3 P) + \frac{\Lambda(t)}{3}, \ee
and the continuity equation derived from \eqref{inject} is
\be\label{eq33}
 {\rho_\rad^\prime}+ 3 \frac{a^\prime}{a} (\rho_\rad + P_\rad)  = - \frac{{\Lambda^\prime}(t)}{8 \pi G}.
\ee
Note that equation \eqref{eq33} encodes the diffusion of energy between the dark sector and the energy density of matter\footnote{Our model can be seen as a special case of the so-called interacting dark energy models some of which have been shown to present instabilities under perturbations \cite{Valiviita:2008iv}. It is important to point out that the type of model we consider is free of such pathological behaviour (for details see \cite{deCesare:2021wmk}).}.  The only assumption in the previous equation is that the diffusion process does not disrupt the homogeneity and isotropy\footnote{Thermal equilibrium for the radiation is not necessarily valid in our context. Even when this is true during a certain regime of the high energy/density primordial universe, thermal equilibrium might be hard to maintain as we approach Planck scales (if we take seriously the extrapolation of particle physics to those scales). The reason is that the condition  for thermal equilibrium $\Gamma>H$ on the interaction rate $\Gamma$ (where $\Gamma\equiv n \sigma$ where $n\sim T^3$ is the number density and $\sigma$ the cross section for interactions) cannot continue to hold close to $T_p=m_p$ because $\sigma\sim 1/T^2$ for high energy processes and thus $\Gamma\sim T$. As $T\propto \rho_{\rm rad}^{1/4}$ drops dramatically if the initial $\rho_0$ is in thermal equilibrium (see Figure \ref{fig:loglog}), while $H$ remains close to the Planck scale $H\approx m_p>T$, all species decouple in the inflationary past and the radiation injection via the decaying $\Lambda$ cannot achieve thermal equilibrium until later when $\rho_{\rm rad}$ eventually grows larger than $H^4$.}  of the background matter and geometry configurations to leading order (perturbations will be considered but they will be small in comparison with average densities). As stated before, we assume that the relevant channel into which $\Lambda$ decays is massless fields so that $\rho_\rad=3P_\rad$ (this justifies the subindex `$\rad$' in our notation, $\rho$ and $P$ denoting the total energy density and pressure that will have contributions coming from the Higgs scalar). 
\vskip.3cm

In such case equations (\ref{eq1}) and \eqref{222} can be combined to obtain
\begin{equation}\label{eq2}
\begin{aligned}
  {a^{\prime\prime}} + 4 \frac{a^{\prime 2}}{a} = \frac{ 2}{3 a^{5}} \Lambda( t ),
   \end{aligned}
\end{equation}
which directly relates the dynamical behaviour of $\Lambda(t)$ and the scale factor. 
If the initial cosmological constant \eqref{lala} starts at its natural Planckian value $\Lambda(0)=\Lambda_0\sim m_p^2$, the initial conditions for the matter density are not important. Indeed, the dynamics of the initial phase of the cosmic evolution is basically insensitive to the value of $\rho(0)$ in the range $0\le \rho(0)\le m_p^4$ (this is a standard aspect of the the usually emphasized robustness of inflation: matter density decays exponentially during the De Sitter phase and becomes rapidly irrelevant for the background evolution).  As we show below, our model shares this property with standard inflation as long as $\beta$ is sufficiently small.

Notice that the Hubble rate in terms of 4-volume time is
\be\label{H}
H\equiv \frac{\dot a}{a}=\frac{a^3 a^\prime}{a}=\frac13 \frac{d a^3}{dt},
\ee
where the symbol $\cdot$ denotes derivatives with respect to comoving time $\tau$.

 During the initial phase of expansion defined by the condition $\beta m_p t <1$ the universe behaves  approximately as a de Sitter universe with Hubble rate $H=H_0\approx\sqrt{\Lambda_0/3}\sim m_p$ (there are small corrections to the Hubble rate due to the rest of matter fields that are not important here; we will come back to them in what follows).  During that initial phase we can integrate  the previous equation (with the initial condition $a(0)=1$) and find that $a^3\approx  3H_0 t+1$. Equivalently, during such period equation \eqref{lala} can be rewritten as \be \label{endy}\Lambda(a)\approx \Lambda_0 \exp(-\frac{m_p}{3H_0}\beta (a^3-1)), \ee which yields an extremely flat curve---seen in Figure \ref{fig:loglog}---for 
\be \label{condimentos} \beta a^3<3 \frac{H_0}{m_p},\ee with a sharp descent for $\beta a^3\approx 3 H_0/m_p$ (assuming $\beta\ll 1$).  This implies that for sufficiently small $\beta$ the background evolution will be very similar to that of standard inflationary cosmology. 
For instance one can get the inflationary phase to last for about 60 e-folds (order of magnitude of what is needed, see Section \ref{efo}), $\sN\equiv \log(a_{\rm end}) \sim60$ if $\beta$ is  
\be
\beta \approx \frac{3 H_0}{m_p} 10^{-80}.
\ee These estimates are confirmed by the numerical solution of the previous equations illustrated in Figure \ref{fig:loglog}. 

\subsection{Higgs Dynamics during the De Sitter phase}\label{higgies}

In the model that we are presenting the background dynamics is dominated by the relaxing cosmological constant. In such framework there is no need for the 
inflaton field of standard inflationary scenarios. Nevertheless,  a (single) scalar field degree of freedom is still necessary for the mechanism of structure formation proposed here to work in its simplest form: this allows for the use of certain conservation laws for super-Hubble modes allowing us to predict the amplitude of perturbations at the CMB from initial condition during inflation (this is sometimes referred to as the Weinberg theorem \cite{Weinberg:2008zzc} whose proof we revisit in the Appendix \ref{WT}).  In addition---as the source of structure will be the hypothetical fundamental Planckian granularity, and, as mentioned in the introduction---the degree of freedom interacting with such fundamental inhomogeneities at the Planck scale must be scale-invariance-breaking in nature. Therefore,   the Higgs field is the natural carrier of the inhomogeneities as, on the one hand, it is the single scalar degree of freedom in the standard model of particle physics, and, on the other hand it is the mediator of the breaking of scale invariance. Even when it is quite possible that a different realization of our scenario might exist, we will concentrate here on such minimalistic model where only the physics of the standard model enters into consideration as far as the description of matter is concerned.  It is also important to point out that the necessity of a scalar field degree of freedom 
is rooted only in its role in the mechanism of structure formation (described in detail in the following section) as the Higgs here plays no important role in the dynamics of the background.  For that reason, our model should not be confused with models of Higgs inflation \cite{Bezrukov:2010jz, Bezrukov:2007ep, Sloan:2018osd}. 

However, the dynamics of the Higgs during the inflationary era will be central in the model so we review it in detail here.
The Higgs field equation in the FLRW background is
\be\label{zeromy}
\ddot \phi_0+3 H_0 \dot\phi_0
%+\Gamma_\phi \dot\phi_0 
+ \Gamma_{\rm Planck} \dot\phi_0+ \frac{d V(\phi_0)}{d \phi} =0.
\ee where the term $\Gamma_{\rm Planck} \dot\phi_0$ is a friction term associated to the energy loss caused by the production of inhomogeneities mechanism that we will introduce in Section \ref{sf}. There we will see that $\Gamma_{\rm Planck}  \ll H_0$ and thus this term can be  safely neglected  from the previous equation when analyzing the dynamics of the zero mode of the Higgs\footnote{ Strictly speaking, another term  $\Gamma_\phi \dot\phi_0$ encoding a standard form of diffusion representing the particles generated via the interactions of the Higgs with the rest of the fields in the standard model should be added to equation \eqref{zeromy}.  The quantity $\Gamma_\phi$ is determined by the known interactions of the standard model (which we are assuming here to make sense all the way to close to the Planck scale).  It follows from the interaction structure of the standard model that the decay rate $\Gamma_\phi$ must be quadratic in the relevant couplings  times some energy scale. Taking this energy scale to be in the natural scale, i.e. $H_0$,  we get $\Gamma_\phi\approx \alpha H_0$ for some dimensionless constant \cite{Albrecht:1982mp}. It can be argued that $\alpha\ll 1$ and thus this term is negligible  in our case.}. 
We assume that $\phi_0$ is in the usual `terminal velocity' configuration where $3H_0 \dot\phi_0=-\partial_{\phi} V[\phi_0]$
which, from $V\approx (\lambda/2) \phi^4$ (where $\lambda$ is the self-coupling constant of the Higgs) implies\footnote{One can explore this numerically and for initial `velocities' away from \eqref{doty} there is a transition time where (as expected) the terminal velocity approximation is not valid. However, even when starting from the (large) natural Planckian value dictated by dimensional analysis $\dot \phi_0\approx -m_p^2$ the scale factor enters into the terminal velocity regime after a few e-folds when the Higgs scalar starts rolling back  towards the Planck scale. 
%(i.e. $\dot \phi_0>0$). 
We are assuming that the Higgs quartic term dominates the potential for values $m_{\rm H}\ll \phi_0\lesssim H_0\approx m_p$. We also treat the Higgs as a single scalar field (for presentation simplicity) ignoring in our equations its ${\rm su}(2)$ internal indices. Our expressions make sense in a polar decomposition $\phi^A=\phi\, v^A$ with $v^A \in {\rm su}(2)$ and $v^Av_A=1$.}
\be\label{doty}
\dot\phi_0\approx-{2\lambda} \frac{\phi_0^3}{3 H_0 }.\ee
Here we are using that $\Gamma_{\rm Planck}/H_0 \ll 1$ (as mentioned, this assumption will be shown to be valid later when we derive equation \eqref{gamita}).
We will assume that the Higgs starts with a large expectation value 
\be \label{ef}
\phi_0(0)\approx m_p.
\ee 
From \eqref{doty} we get $|\dot \phi_0|\ll H_0^2$ as long as $|\lambda|\ll 1$.
The requirement that the universe is dominated by the cosmological constant, namely $|V[\phi_0]|=(|\lambda|/2) \phi_0^4 \ll \Lambda m^2_p/(8\pi)=H_0^2m_p^2/(8\pi)$ is automatically satisfied if $|\lambda| \ll 1$. For these reasons, one can neglect the effects of the potential in the dynamics of the background geometry in the De Sitter phase studied in Section \ref{desi}.   Finally,
in the terminal velocity regime we have (from the time derivative of \eqref{doty}) that
\be
\ddot \phi_0\approx \frac{12}9\lambda^2 H_0^3,
\ee  
which will be neglected as a higher $\lambda$ correction in the perturbation theory calculations that follow.
The picture in Figure \ref{fig:loglog} will have to be modified when $\Lambda(t)\approx V[\phi_0]/m_p^2$. This happens after the end of inflation, and thus away from the region where the seed of structure formation are produced as discussed in Section   \ref{sf}. 

From \eqref{doty} one finds  solutions
\be\label{higgy}
\phi_0(\tau)=\frac{m_p }{\sqrt{1+ \frac 43 \frac{m_p^2}{H_0} 
   \lambda  \tau }}\ \ \ \ {\rm or }\ \ \ \ \phi_0(a)=\frac{m_p }{\sqrt{1+ \frac 43 \frac{m_p^2}{H^2_0} \lambda \log(a) }}.
\ee
Note that during the first e-folds, say $\sN=\log(a)\sim 11$ (which approximately correspond to the period during which the fluctuations visible in the CMB are produced in our model),  and for $\lambda\approx -10^{-2}$ (which is the correct order of magnitude value in the standard model close to the Planckian scale \cite{Degrassi:2012ry}) the Higgs changes slowly for $H_0\approx m_p$ as assumed, namely 
\be\label{47}
\frac{\Delta\phi_0}{\phi_0}\lesssim 10^{-1}.
\ee

\subsection{Radiation generated by the decaying $\Lambda$ }

The analysis of the background dynamics, given in Section \ref{desi}, relies on neglecting the effect of the radiation emitted as $\Lambda$ decays in matter field modes. In addition, there is the question of how the initial conditions for radiation affect the conclusion of Section \ref{desi}. Here we show that none of these neglected aspects have an important influence and that results of the previous simplified analysis remain correct to the level of approximation considered. The physical reason is that the cosmological constant term dominates the Friedmann equation due to its slow decay in $a$ while the radiation dilutes as $a^{-4}$ as the energy injection \eqref{eq33} for $\beta\ll 1$ is negligible at first.  Eventually, energy injection becomes comparable with the dilution rate and radiation density $\rho_\rad$ starts growing again.  One can understand these features---which where first exibited by the numerical solution of the equations as plotted in Figure \ref{fig:loglog}---semi-analytically giving a closer look at equation \eqref{eq33} which for diffusion into radiation becomes \be \frac{d\rho_\rad}{da}+\frac{4}{a}\rho_\rad   = - \frac{{\Lambda^\prime}(t)}{8 \pi G  a^\prime}
%+\frac{\Gamma_\phi \dot \phi_0^2}{\dot a}
.\ee
Using that $H=\sqrt{\Lambda/3}=a^2  a^\prime\approx H_0= \sqrt{\Lambda_0/3}$ and also $\dot a\approx H_0 a$---recall equation \eqref{H}---we get
\be\label{bons-enfants} \frac{d\rho_\rad}{da}+\frac{4}{a}\rho_\rad   \approx \frac{3 \beta a^2}{8\pi}m_p^4
%+ \frac{4{\lambda^2} \epsilon_\phi^6 \alpha^{\star} } {a (3+\alpha^{\star})^2} { H_0^4}
,\ee
where we have used equation \eqref{doty} and the fact that $\phi_0\approx\phi_0(0)\approx H_0\sim m_p$. Integrating \eqref{bons-enfants} we obtain
\be\label{rodri}
\rho_\rad\approx\frac{\rho_0 }{a^4}+\frac{3 \beta a^3 }{56 \pi} m_p^4 .
%+ \frac{{\lambda^2} \epsilon_\phi^6  \alpha^{\star} } {(3+\alpha^{\star})^2} { H_0^4}
\ee
%which shows that the injection of energy from the Higgs decay (the last constant contribution) is completely negligible for the values of $\alpha^{\star}$ corresponding to the standard model in comparison with the energy density scale $H_0^2 m_p^2$ from the cosmological constant (recall $\lambda\approx 10^{-2}$). 
Therefore, our first approximation \eqref{eq33} turns out to be fine. Thus,  we simply ignore that last constant contribution to the radiation density in the previous equation in order to simplify the presentation. However, a similar 
constant density contribution coming from diffusion will play a key role in the discussion of Section \ref{de}.
Notice that the minimum in the radiation density observed in Figure \ref{fig:loglog} can be estimated from the condition $d\rho_\rad/da=0$ and gives 
\ba a_{\rm min}^7 &\approx& \frac{224 \pi  \rho_0 }{9 \beta   m_p^4}\n \\ 
\rho_{\rm min}&\approx&\frac{7\rho_{0}}{3 a_{\rm min}^4},
\ea with $a_{\rm min}\approx e^{27}$ and $\log(\rho_{\rm min}/m_p^4)\approx -107$ for the parameters in Figure \ref{fig:loglog}. With a bit of abuse of the approximation we can estimate the radiation at the end of the inflationary period when $\beta a^3_{\rm end}=3H_0/m_p$, as implied by  (\ref{endy}).  One gets \be \label{fine}\rho_{\rm end}\approx \frac{9 m_p^4}{56 \pi}\equiv T_{\rm end}^4.\ee We see that the previous semi-analytic argument reproduces well the qualitative features of the numerical solution in Figure \ref{fig:loglog}. Notice that the final `reheating temperature' $\approx \rho_{\rm end}^{1/4}$ is independent of the initial conditions and of the order of Planck temperature. 

\subsection{Estimate of the lifetime of $\Lambda$ after the inflationary era and number of e-folds}\label{efo}

The numerical evolution shows that soon after we reach the end of inflation the universe becomes quickly dominated by radiation with an initial radiation density which is estimated from \eqref{rodri}. The end of inflation is characterised here by the condition
\be\label{condimento}
\beta a_{\rm end}^3\approx 3 \frac{H_0}{m_p},
\ee
which follows from equation \eqref{endy}.
The Friedmann equation \eqref{eq1} in the radiation dominated domain becomes
\be
a^4 a^\prime=\sqrt{\frac{8\pi \rho_{\rm end}}{3m_p^2}} a^2_{\rm end}.
\ee
Integrating and  multiplying by $\beta m_p$ we get
\be
\frac{1}{5} \sqrt{\frac{ 3m_p^4} {8\pi \rho_{\rm end}}} \beta a^3_{\rm end} \left(\left(\frac{a}{a_{\rm end}}\right)^5-1\right)=\beta m_p \Delta t.
\ee
Neglecting the $-1$ inside the parenthesis, using \eqref{condimento} to eliminate the $\beta$ dependence,  replacing for $\rho_{\rm end}$ using \eqref{fine}, and assuming that $a/a_{\rm end}\approx T_{\rm end}/T$,  we obtain the following expression for the dependence of $\Lambda$ on temperature after inflation
\be\label{decayy}
\Lambda=\Lambda_0 \exp\left(-\beta m_p \Delta t\right)\approx \Lambda_0 \exp\left[ -\frac{\sqrt{21}}{5} \left(\frac{9}{56 \pi}\right)^{\frac{5}{4}}\left(\frac{H_0}{m_p}\right)^{\frac{5}{2}} \left(\frac{m_p}{T}\right)^5\right],\ee
   which implies that the cosmological constant becomes negligible in comparison to the present value $\Lambda_{\rm today}=10^{-120} m_p^2$ extremely quickly by the time when  the temperature of the universe is still close to Planckian. The important point here is that this is well before the electro weak transition temperature $T_{\rm ew}\approx 10^{-17} m_p$ so that the essentially vanishing cosmological constant can grow again via the mechanism presented in \cite{Perez:2017krv, Perez:2018wlo} to the present observed value.  

The parameter $\beta$ chosen in the Figure \ref{fig:loglog} corresponds to an illustrative value. Here we analyze in more details observational constraints on this value. 
An important feature of our model is the generation of inhomogeneities in an approximately scale invariant fashion as observed on the CMB during the quasi De Sitter phase. The scale of these fluctuations range from $L_{\rm min}=10^{-2} {\rm Mpc}$ to $L_{\rm max}=10^3 {\rm Mpc}$ today. Even when the mechanism for structure formation will be different from the inflaton `vacuum fluctuations' of the standard paradigm, the De Sitter regime of inflation will still play a key role. In particular, one needs the scales of fluctuations visible today to correspond to the Hubble scale $H\approx H_0\sim m_p$ at the time of inflation. This demands a minimum number of e-folds from the beginning of inflation to today
\be
\sN^{{\rm start}\to {\rm today}}_{\rm min}=\log(H_0 L_{\rm max})=\log\left(\frac {H_0} {m_p}\right)+\log(m_p L_{\rm max})=\log\left(\frac {H_0} {m_p}\right)+138.
\ee  
On the other hand the number of e-folds since the end of inflation $N^{end\to {\rm today}}_{\rm min}$ is
\be
\sN^{{\rm end}\to {\rm today}}_{\rm min}=\log(T_{\rm end}T_0^{-1})\approx \frac{1}{4}\log\left(\frac{9 H^2_0}{56m^2_p \pi}\right)+74,
\ee
from which we get necessary minimum number of inflationary e-folds 
\be\label{postilla}
{\sN^{{\rm start}\to {\rm end}}_{\rm min}\approx \frac{1}{2}\log\left(\frac {H_0} {m_p}\right)+65.}
\ee

\section{Structure formation}\label{sf}

As described in previous sections, the cosmological constant decays spontaneously due to diffusion into radiation degrees of freedom exponentially in unimodular time as a result of quantum gravity instability associated to the fundamental granularity. 
We have shown that this produces a quasi De Sitter dynamical evolution for the universe. We have also assumed that the Higgs potential starts in a homogeneous configuration with a natural expectation value $\phi_0\sim m_p$ and have shown how $\phi_0$ is expected to evolve. Even when negligible in such a dynamics (as it will be shown later) we included a friction term controlled by  $\Gamma_{\rm Planck}$ in \eqref{zeromy}. This term is produced, we argue, by the interaction of the Higgs scalar with the Physics at the fundamental scale: the Planckian granularity. This interaction of the homogeneous Higgs $\phi_0$ and the inhomogeneous granular structure at the Planck scale---mediated by the scale-invariance-breaking of the Higgs---will generate (or excite) inhomogeneities in $\phi$ that are born at the Planck scale via a stochastic process described in detail in Section \ref{pan}).   

Without knowing the dynamics of the deep quantum gravity regime it is hard to construct a fundamental account for the interaction between the granular Planckian structure and the matter fields involved. Nevertheless, guided by dimensional analysis and expected features of quantum gravity when considering its compatibility with low energy Lorentz invariance we will construct a model with very few free parameters. This program has certainly an important degree of speculation. However, there are known instances in physics where general conceptual reasoning together with dimensional analysis can lead to meaningful insights about a physics that might be, at first, hard to describe in fundamental terms. This is the perspective we adopt in this section.
  
During the initial De Sitter phase the scalar curvature is close to Planckian so that the scale of discreteness could naturally catalyse the emergence of inhomogeneities.  As argued here, and in \cite{Perez:2017krv}, the discreteness scale should play a role in those field theoretical degrees of freedom which are not scale invariant. These are the degrees of freedom that, from a relational perspective, carry a `ruler'   or  `reference frame' with respect to which the fundamental quantum gravity scale $\ell_p$ can become meaningful. In this sense it is natural to accept that as a result of such interaction inhomogeneities should be created in the Higgs scalar (which is the degree of freedom that introduces the breaking of scale invariance in the standard model).  
The energy flow involved in this can be parametrized (phenomenologically) as an Ohmian diffusion term in the equation of motion of $\phi_0$.  As this effect is assumed to have a quantum geometry origin, and as the only relevant geometric scale around is the Hubble rate, dimensional analysis suggests the diffusion to be characterized by  a dimensionless coefficient $\gamma$ is a dimensionless coefficient as follows (note that the energy density change per unit time encoded in the l.h.s. is measures in $(energy)^5$ units)   
\be\label{clavisima}
\Gamma_{\rm Planck}\dot \phi_0^2=\gamma  H^5,
\ee
where the previous is the diffusion term in  \eqref{zeromy} and we assume  $\gamma\ll 1$ (this assumption will be confirmed by the analysis that follows). 
Such friction term induces an additional steady contribution to the divergence of the Higgs energy momentum tensor component which will be absorbed by the generated inhomogeneities (quantitatively this will be described by suitable continuity equations written below).
We will see below that, such steady  injection of energy into the fluctuations (via the mechanism evoked in this paragraph but made mathematically precise below) produces a spectrum of scalar perturbations in the Higgs that is adiabatic and approximately scale invariant.
We will also show that---using the Weinberg theorem to analyze the effect of these at CMB times---the magnitude of the parameter $\gamma$ needs to be fixed to $\gamma\approx 10^{-16}$ which is, remarkably,  of the same order of magnitude as the dimensionless number $\gamma_{\rm H}\approx 10^{-17}$ charaterizing the breaking of scale invariance by the Higgs, recall \eqref{conficinfi}.

Note that from the semiclassical perspective of quantum field theory on curved spacetimes we must also note that there are no ambiguities in the notion of particles for conformal invariant  quantum fields as the FLRW background is conformally flat. As a result there is no real particle creation for such modes if thought of as test fields on the cosmological conformally flat background \cite{Birrell:1982ix} \footnote{The natural state for conformally invariant fields in the De Sitter phase is the Bunch-Davies vacuum (any deviations from it are exponentially diluted during inflation). This state coincides with the Gibbons-Hawking state that is perceived as a thermal state with temperature $T_{\rm gh}=H_0/(2\pi)$ by any freely falling observer \cite{Gibbons:1977mu}. However, such a thermal bath should be regarded as the analog of the Unruh particles in flat spacetime. They are there but have an elusive physical reality as can be clearly seem by considering the examples of a spacetime that is initially flat, then De Sitter spacetime in an intermediate region, and finally flat again.  If one starts with the Poincare vacuum state then the state will evolve into something well approximated by the Bunch-Davies vacuum in the intermediate phase (with Gibbons-Hawking temperature $T_{\rm gh}$). However, the state will emerge in the final state as the poincare vacuum again. No real particles are created by the De Sitter phase. Therefore, such `thermal excitations' due to the presence of the De Sitter horizon in the initial phase of evolution in our model, cannot be responsible for the real fluctuations that we need to find in the future stage where the universe has gone out of the De Sitter phase and the horizon has become virtually infinite (like in Minkowski).}. For degrees of freedom breaking scale invariance the situation is the same as long as we concentrate on scales well within the Hubble radius. However, the notion of particle (and their number) becomes ambiguous as soon as we consider modes with super Hubble wavelength. The mechanism of excitation of inhomogeneities discussed above is producing particles at around the scale where the notion of particles become ambiguous. By this we are not saying that a complete semiclassical description is at all possible (as any fundamental explanation of the role of discreteness would need to appeal to quantum gravity). Nevertheless, we are arguing here that the excitation of inhomogeneities that we postulate is taking place close to the scale where the semiclassical account allows for something peculiar to happen. 

The peculiar physical aspect that we are evoking is rooted in the UV structure of our physical description of matter and geometry. In this respect it is important to recall the discussion of the renormalization of the energy momentum tensor in quantum field theory on curved spacetimes, and the fact that UV contributions lead to an (anomalous from the pure quantum field theory perspective) violation of energy momentum conservation (equation \eqref{yep}). We will see that in the case of the Higgs scalar such anomaly could be interpreted as the source of the term $\Gamma_{\rm Planck}\dot \phi_0$ in \eqref{zeromy} (or its possible semiclassical description). We will come back to this in the discussion section once the implications of the present perspective are spelled out.

\subsection{Phenomenological analysis}\label{pan}
Let us start from the study of the dynamical equation for the scalar field inhomogeneities.
In standard treatments  these perturbations are quantized and assumed to be in some (preferred) vacuum state (say the Bunch-Davies vacuum).
The inhomogeneities that we see today in the CMB are assumed to arise from the quantum fluctuations which somehow become classical by the time that they leave their imprint on the visible sky. Although such perspective is largely adopted in the community,  there no complete consensus on its internal consistency. 
We have already mentioned the well known trans-Planckian tension. On another front it has been noted that it suffers from intrinsic interpretational problems associated with the measurement problem in quantum mechanics of a close system (see  \cite{Perez:2005gh} for a discussion of these issues and further references).

Motivation for our model is fuelled in part by these conceptual tensions which we try to alleviate by proposing an alternative. Notice that in contrast to the standard account, inhomogeneities in the model that we propose here arise from an actual physical interaction that actively produces inhomogeneities  
on the background Higgs  value $\phi_0$. Even when these fluctuations and the background field configurations are intrinsically quantum, 
we will represent them by semiclassical states whose expectation values are assumed to be well approximated by classical field equations. 

Consequently, we define perturbations of the zero mode $\delta\phi_k$ with wave number $k$ for which the following field equation holds
\be\label{km}
 \delta \ddot\phi_k+3 H_0 \delta\dot\phi_k+\frac{k^2}{a^2} \delta\phi_k+\frac{d^2V(\phi_0)}{d \phi^2} \delta\phi_k=0.
\ee
The last term in \eqref{km} is smaller that the third term when $k=a H_0$ (i.e. at horizon crossing)  and it is in general suppressed by the Higgs self coupling so we will treat its influence in perturbation theory below. This follows from
\be\label{penrose}\frac{d^2V(\phi_0)}{d \phi^2} = 6\lambda \phi_0^2 \ll H_0^2,\ee which would automatically hold for $\phi_0\sim H_0$ as $\lambda\ll 1$ in this large field regime.
Given these assumptions, and according to  \eqref{km}, super-Hubble modes (for which $k\ll aH_0$) satisfy (to zeroth order in $\lambda$)
\be\label{death}
 \delta \ddot\phi_k+3 H_0 \delta\dot\phi_k\approx 0 \ \ \ \ {\rm or\ equivalently}\ \ \ \ \frac{d (a^3\dot \delta\phi_k)}{dt}\approx 0,
\ee
which implies
\be\label{nobep}
\delta\phi_k(\tau)=q_k \frac{ e^{-3 H_0 \tau}}{3 H_0}+\delta\phi_k \ \ \ \ {\rm or\ equivalently}\ \ \ \ \delta\phi_k(\tau)=\delta\phi_k+\sO(a^{-3}) \ \ \ \ {\delta\dot \phi_k}=\sO(a^{-3}),
\ee
for some $q_k$.
Super-horizon modes freeze out and their time derivative $\delta\dot\phi_k$ decays exponential in co-moving time or as $a^{-3}$ in terms of the scale factor \footnote{If we keep the contribution of the potential in equation \eqref{km} then one gets instead that \be \frac{\delta\dot \phi_k(\infty)}{\delta\phi_k(\infty)}=-\frac32 \left(1-\sqrt{1-\frac83\lambda}\right)\approx -2\lambda. \ee The fact that $\lambda$ is negative (Higgs instability) introduces a growing mode that goes like
$\approx \exp(-4\lambda H_0 \tau)$. However, for $\lambda\approx -10^{-2}$ this growth is sufficiently slow to grant the validity of perturbation theory until the end of inflation $H_0\tau\approx 60$.}.  

\subsubsection{Energy fluctuations and the power spectrum}

The effect of the Planckian granularity will be modelled by a Brownian diffusion process that injects energy in the Higgs scalar by leaving an imprint of the fundamental scale as inhomogeneities in the background value. More precisely,  we consider a stochastic process generating density fluctuations by excitation of the Higgs scalar modes at the Planck scale which is assumed to coincide with the curvature scale $H_0\approx m_p$. As in the description of the Brownian motion, the process  stochasticity is an assumption that allows for a statistical effective description of the effect of a large number of underlying independent microscopic degrees of freedom whose individual dynamics can be understood only in terms of a (more fundamental) quantum gravity analysis.

The caracterization of the stochastic process  requires the analysis of the energy cost of generating the inhomogeneities in the Higgs. For that purpose   
let us first write the Higgs scalar as $\phi(x)=\phi_0+\delta \phi(x)$ so that  the first order perturbation of the energy density (up to second order) is  
\ba\label{ppp}
\delta\rho\equiv \delta T_{00}((x^\mu))&=& \frac{1}{2} \delta \dot \phi^2+\frac{1}{2a^2} \delta \vec\nabla\phi^2+\delta V(\phi)
\\&\approx&\dot \phi_0 \delta\dot\phi(x^\mu)+\frac{d V(\phi_0)}{d\phi}\delta\phi(x^\mu)+\frac{1}{2}\delta\dot\phi(x^\mu)^2+\frac1{2 a^2}(\vec \nabla\delta\phi(x^\mu))^2+\frac{1}{2} \frac{d^2 V(\phi_0)}{d\phi^2}\delta\phi(x^\mu)^2.\n
\ea
At this point, it is important to point out two important features of the previous expression. First the perturbation of the energy momentum tensor in equation \eqref{ppp} is obtained by assuming that the Higgs is a test field (i.e. metric perturbations are excluded here). Second, we have expanded up to second order in perturbation while in the usual cosmological 
perturbation theory one only needs to go up to first order. We will see below that in all dynamical considerations involving gravity we will restrict to linear perturbations. Very importantly, during the De Sitter phase, scalar metric perturbations turn out to be trivial (see equation \eqref{desitete} below) which implies that (at least during that period) the test field energy momentum tensor and the full linearized energy momentum tensor actually coincide. This is not the case for the second order perturbations. However, the later will only be used as an interpretational devise that offers the means to talk about energy flows involved in the creation of the perturbations by the stochastic process that describes the interaction between the discreteness scale and the Higgs scale \footnote{ This is analogous to the discussion of energy flow in Hawking black hole radiation where back reaction is neglected and the field degrees of freedom are considered those of a test field. However, reliable physical information is captured by such notion allowing for the clear understanding of the physical consequence of particle creation ranging form negative energy flows across the horizon, to the violations of the classical area law, and the energy loss via evaporation at infinity.}.

We assume that $\delta \phi(x^\mu)$ is a stochastic variable with a probability distribution such that the associated linear momentum vanishes, namely 
 \be\label{stochy} \sbraket{\delta\phi(x^\mu)}=0,\ee
 where from now on $\sbraket{\ }$ denote ensemble averages which are to be distinguished from quantum expectation values $\braket{\ }$.
It follows  that 
\be\label{nina}
\sbraket{\delta T_{00}(x^\mu)}\approx \frac{1}{2}\sbraket{\delta\dot\phi(x^\mu)^2}+\frac1{2a^2}\sbraket{(\vec \nabla\delta\phi(x^\mu))^2}+\frac{1}{2} \frac{d^2 V(\phi_0)}{d \phi^2} \sbraket{\delta\phi(x^\mu)^2}, 
\ee
where the approximate sign comes from the fact that we have truncated the expansion of $V(\phi)$ to second order in $\delta\phi(x^\mu)$.
The previous equation implies that the (ensemble) average energy contribution to the field perturbations in the stochastic process is controlled by the second order terms in the expansion (to leading order). 
We can relate the second moments of the probability distribution to the power spectrum of perturbations if we decompose the field in Fourier components
\be
\delta\phi(t, \vec x)=\frac{1}{(2\pi)^{\frac32}} \int dk^3 \delta \phi_{\vec k}(t) \exp(i \vec k\cdot \vec x),
\ee
with reality conditions
\be
\overline{\delta \phi_{\vec k}}=\delta \phi_{-\vec k}.
\ee
The standard definition of the 2-point correlation function (see for instance \cite{Peter:1208401, Brandenberger:2003vk}) is defined by
\ba
\xi_{\phi}(\vec r)&\equiv& \sbraket{\delta\phi(\vec x)\delta\phi(\vec x+\vec r)} \n \\
&=&\frac{1}{(2\pi)^3} \int dk^3 dq^3  \sbraket{\delta \phi_{\vec k}\delta \phi_{\vec q}} \exp(i (\vec k+\vec q)\cdot \vec x),
\ea
where the second line has been expressed in terms of Fourier modes.
Due to the background symmetries the stochastic process creating the perturbations the 2-point correlation function must be
homogeneous (independent of $\vec x$) and isotropic. 
In terms of Fourier modes this implies that 
\be\label{ps}
\sbraket{\delta \phi_{\vec k}\delta \phi_{\vec q}} =P_{\delta\phi}(k) \delta^{(3)}(\vec k+\vec q), 
\ee
where $P_{\delta \phi(k)}$ is the power spectrum of the perturbations $\delta\phi$. 
The previous equation implies the key relationship between the power spectrum and the 
expectation value of the square of the perturbation at the same point $\delta\phi(\vec x)$, namely
\be\label{keyly}
\sbraket{\delta\phi(\vec x)\delta\phi(\vec x)}=\frac{1}{(2\pi)^3} \int dk^3 P_{\delta \phi} (k). 
\ee
This relation allows, as we shall show below, to express the ensemble average of the energy momentum tensor 
of the perturbations in terms of their power spectra.

\subsubsection{Energy momentum conservation (continuity equations)}\label{conti}

In this section we study the equation of state of the Higgs perturbations when averaged in the ensemble representing the probability distribution of the stochastic process, whose general properties were introduced in the previous section. This equation of state will play a role in the continuity equations for the perturbations which will allow us to interpret the energetics of the generation of inhomogeneities. 

In order to write the equation of state we need to consider the ensemble expectation value of the energy momentum tensor. 
Repeating the exercise that led to \eqref{nina}, but now for all the components of the energy momentum tensor, we obtain
\ba\label{noname}
\sbraket{T_{ab}}&=&-\frac{\Lambda m_p^2}{8\pi} g_{ab}+ \sbraket{\nabla_a (\phi_0+\delta\phi) \nabla_b(\phi_0+\delta\phi)-\frac{1}{2}g_{ab} \left(\nabla_c(\phi_0+\delta\phi) \nabla^c(\phi_0+\delta\phi)+2 V((\phi_0+\delta\phi))\right)}\n\\
&=& T^{(0)}_{ab}+ \underbrace{\sbraket{\nabla_a \delta\phi \nabla_b\delta\phi}-\frac{1}{2}g_{ab} \left(\sbraket{\nabla_\alpha\delta\phi \nabla^\alpha\delta\phi}+\frac{d^2 V(\phi_0)}{d \phi^2}  \sbraket{\delta\phi^2 }\right)}_{\sbraket{\delta T_{ab}}},
\ea
where we have expanded to second order in the perturbation and the first order terms are gone due to \eqref{stochy}. The second order terms are (as the zeroth order ones) of the perfect fluid form,  $\sbraket{T_{ab}}=\rho_{\rm h} u_au_b+P_{\rm h} h_{ab}$, where $u^a$ and $h_{ab}$ are the comoving four-velocity and spatial metric of the background respectively.  This follows from the assumption that the stochastic process generating the perturbations is isotropic and homogeneous.  It follows that 
%\ba \label{pipi}\rho &=&\frac{\Lambda m_p^2}{8\pi}+\frac{\dot \phi^2_0}2+V(\phi_0)+\sbraket{\delta\rho^{(2)}}+\rho_{\rm rad},\n \\  P&=&-\frac{\Lambda m_p^2}{8\pi}+\frac{\dot \phi^2_0}2-V(\phi_0)+\sbraket{\delta P^{(2)}}+\frac{\rho_{\rm rad}}{3},\ea
\ba \label{pipi}\rho_{\rm h} &=&\frac{\dot \phi^2_0}2+V(\phi_0)+\sbraket{\delta\rho^{(2)}},\n \\  P_{\rm h}&=&\frac{\dot \phi^2_0}2-V(\phi_0)+\sbraket{\delta P^{(2)}},\ea
where $P_{\rm h}$ and $\rho_{\rm h}$ denote the pressure and density contributions of the Higgs scalar, and the supra index $(2)$ on the right hand side expresses the fact that these terms come from quadratic contributions in the field perturbations. 
In order to get the explicit form of $\sbraket{\delta \rho^{(2)}}$ and $\sbraket{\delta P^{(2)}}$ we observe that
\be
\sbraket{\nabla_a \delta\phi \nabla_b\delta\phi}=\sbraket{\delta\dot\phi^2} u_au_b+\frac{1}{3 a^2} \sbraket{\vec\nabla\delta\phi\cdot \vec\nabla\delta\phi} h_{ab},
\ee
from which, when replacing back into \eqref{noname},  we get 
\ba
\sbraket{\delta T_{ab}}&=&\sbraket{\nabla_a \delta\phi \nabla_b\delta\phi}-\frac{1}{2}g_{ab} \left(\sbraket{\nabla_\alpha\delta\phi \nabla^\alpha\delta\phi}+\frac{d^2 V(\phi_0)}{d \phi^2}  \sbraket{\delta\phi^2 }\right) \\
\n &=&\frac{\sbraket{\delta\dot\phi^2} +\frac{1}{a^2} \sbraket{(\vec\nabla\delta\phi)^2}+\frac{d^2 V(\phi_0)}{d \phi^2}  \sbraket{\delta\phi^2 }}{2}u_au_b +\frac{\sbraket{\delta\dot\phi^2} -\frac{1}{3a^2} \sbraket{(\vec\nabla\delta\phi)^2}-\frac{d^2 V(\phi_0)}{d \phi^2}  \sbraket{\delta\phi^2 }}{2}h_{ab}, 
\ea
thus, using \eqref{pipi}, we obtain
\ba\label{pdm}
\sbraket{\delta \rho^{(2)}}&=&\frac{\sbraket{\delta\dot\phi^2} +\frac{1}{a^2} \sbraket{(\vec\nabla\delta\phi)^2}+\frac{d^2 V(\phi_0)}{d \phi^2}  \sbraket{\delta\phi^2 }}{2}\approx \frac{1}{2 a^2} \sbraket{(\vec\nabla\delta\phi)^2}+\frac{1}{2}  \frac{d^2 V(\phi_0)}{d \phi^2}  \sbraket{\delta\phi^2 }\\ \label{pdm2}
\sbraket{\delta P^{(2)}}&=&\frac{\sbraket{\delta\dot\phi^2} -\frac{1}{3a^2} \sbraket{(\vec\nabla\delta\phi)^2}-\frac{d^2 V(\phi_0)}{d \phi^2}  \sbraket{\delta\phi^2 }}{2}\approx- \frac{1}{6 a^2} \sbraket{(\vec\nabla\delta\phi)^2}-\frac{1}{2}  \frac{d^2 V(\phi_0)}{d \phi^2}\sbraket{\delta\phi^2 },\n
\ea
where we neglected the $\delta\dot \phi$, as justified by \eqref{nobep} in the $k\ll aH_0$ regime (which is the regime where all these equations will be used).

The previous equations allow for expressing the energy-cost that it would take to create inhomogeneities in the Higgs scalar. Thus, these equations a key in 
describing our stochastic mechanism generating the perturbations. Concretely, one needs to concentrate on the associated continuity equations that encode the energy transfer into inhomogeneities mediated by the stochastic process.
One can think of the Higgs perturbation contribution to the continuity equations as the derivative of work $W^{\rm pert}$ done by the Brownian diffusion on the Higgs with respect to the suitable time parameter. Using the scale factor as such parameter  we get 
\ba
\frac{dW^{\rm pert.}(P_{\delta\phi})}{da} &\equiv& \frac{1}{\dot a}\left(\sbraket{ \delta\dot \rho^{(2)}}+3 H_0 \left(\sbraket{\delta\rho^{(2)}}+\sbraket{\delta P^{(2)}}\right)\right)\n 
\\ &=&\frac{d\sbraket{\delta\rho^{(2)}}}{da}+\frac{3}{a}\left(\sbraket{\delta\rho^{(2)}}+\sbraket{\delta P^{(2)}}\right),
\ea
where we explicitly write the dependence of $W^{\rm pert.}(P_{\delta\phi})$ on the power spectrum of  inhomogeneities $P_{\delta \phi}$ as the work 
depends on squares of the Higgs fluctuations $\delta \phi$ whose ensemble average is encoded, according to equation \eqref{keyly}, in the power spectrum.
The stochastic process will be defined below in more precise terms. However, at this stage we can anticipate that the central equation will be a balance equation of the form 
\be \label{vlowy}
\frac{dW^{\rm pert.}(P_{\delta\phi})}{da}={\rm stochastic \ source},
\ee
thus the remaining tusk is to define the right hand side of the previous equality.
The reason we call this balance equation is that the logic behind the mechanism we propose in what follows is 
similar to the one that leads to Einstein's detail balance equations (an instance of the so-called fluctuation dissipation theorem) in the simple context of Brownian motion.
This analogy is developed further in the following footnote \footnote{ \label{eleven}Consider a free particle moving in a viscous medium. The equation of motion is given by the Langevin equation 
\be
\ddot x+\gamma \dot x=\delta x(t),
\ee
where $\delta x(t)$ is a stochastic noise source representing the fluctuations of $x$ due to the action of the microscopic elements in the environment.
This stochastic variable (which is the analog of the inhomogeities $\delta\phi$ in our cosmological context) has a noise dynamics dictated by a probability distribution such that
\be\label{vlovlo}
\sbraket{\delta x(t)}=0, \ \ \ \sbraket{\delta x(t)\delta x(t')}=P_{\delta x} \delta(t, t'), 
\ee
which are the analog of equations \eqref{stochy} and \eqref{keyly} respectively.
The Langevin equation has a simple solution in the large time asymptotics given by
\be
x(t)=\int _0^te^{-\gamma  t''} \int
   _0^{t''}e^{\gamma  t'}
   \delta x(t')dt'dt''+x_0,
\ee 
from which we get
\be
\dot x(t)=e^{-\gamma t} \int_{0}^t e^{\gamma t'} \delta x(t') dt'
\ee
Consider now the kinetic energy of the particle $E=m \dot x^2/2$, in the long time asymptotic regime conservation of energy requires $\sbraket{\dot E}=0$.  This 
condition can be written using the Langevin equation as
\ba
 \sbraket{\dot E(t)} &=& m \sbraket{\dot x(t) \ddot x(t)}=m \sbraket{\dot x(t) (\delta x(t)-\gamma \dot x(t))}=m \sbraket{\dot x(t) \delta x(t)} -2\gamma\sbraket{E(t)}\n\\
 &=& m e^{-\gamma t}  \int_{0}^t e^{\gamma t'} \sbraket{ \delta x(t') \delta x(t)} dt' -2\gamma\sbraket{E(t)}\n \\&\Longrightarrow & m P_{\delta x}=2\gamma\sbraket{E(t)},
\ea
where we used \eqref{vlovlo} in the last line.
The result is Einstein's detail balance equation relating the expectation value of the energy of the particle (which does not vanish due to the action of the Brownian collisions of the microscopic constituents of the environment), the friction coefficient $\gamma$ (encoding dissipation in the Langevin equation), and the power `spectrum' $P_{\delta x}$ of the fluctuations $\delta x(t)$.  This last equation is the analog of our equation \eqref{vlowy} (whose precise form is derived further down in the paper), the left hand side is in our case a functional of the power spectrum, the right hand side encodes the energy fluctuations produced by the Planckian granularity (precisely defined in equation \eqref{mono}). In the brownian motion case one invokes thermal collisions of the environment on the particle and writes $\sbraket{E}=kT$. In our case we propose that the action of granularity on field fluctuations have the analog effect (see discussion leading to \eqref{mono}).  A key difference with this simple example is that instead of having a single degree of freedom we have an infinite tower of modes that are being excited  as they cross the horizon. This produces a continuous feeding of energy at different modes as the background expands and hence a continuous injection of energy characterized by the formal equation \eqref{vlowy} which will become \eqref{mono} in its concrete realization.  The argument leading to \eqref{mono} is given in the main text.
}.

Replacing \eqref{pdm} and \eqref{pdm2} the previous equation becomes
\ba\label{continuity}
{\frac{dW^{\rm pert.}}{da}\equiv \frac{d\sbraket{\delta\rho^{(2)}}}{da}+\frac{2}{a}\sbraket{\delta\rho^{(2)}}-\frac{1}{a}\frac{d^2 V(\phi_0)}{d \phi^2} \sbraket{\delta\phi^2 } ={\rm stochastic \ source}}.
\ea
Now we are ready to define the right hand side of the previous equation more precisely. The assumption is that the zero mode $\phi_0$, while rolling down the potential and evolving in the spacetime geometry,  interacts with the granularity scale and diffuses energy to the modes with (physical) wave number $k/a\approx m_p$ via the discreteness scale.  
We assume that this work is extracted from the zero mode $\phi_0$ while evolving in the Higgs potential in a stochastically Ohmian way so that its dynamical equation \eqref{zeromy} gets a friction term $\Gamma_{\rm Planck} \dot \phi_0$  
\footnote{
As a simple particular situation illustrating of a rational behind this modification consider a Klein-Gordon scalar field as an example. The field equation $\nabla^a\nabla_a\phi-m^2\phi^2=0$ is explicitly given by 
\ba
\frac{1}{\sqrt{|g|}}\partial_\mu\left(\sqrt{|g|}g^{\mu\nu}\partial_\nu\phi\right)-m^2\phi^2&=&0.\ea
Thus we see from the previous equation that if the background is fluctuating then the equation will get a `Brownian' modification as follows
\be
\nabla^a\nabla_a\phi-m^2\phi^2 =  \xi^a \nabla_a\phi,
\ee
where $\xi^a$ is the contribution from the background fluctuations 
\be
\xi^\nu\equiv -\Delta\left(\frac{1}{\sqrt{|g|}}\partial_\mu(\sqrt{|g|}g^{\mu\nu} )\right).
\ee
In the FLRW context the only non vanishing component of $\xi^a$ allowed by the symmetry is $\xi^0=3\Delta H$---that we called $\xi^0=\Gamma_{\rm Planck}$---is the only possible non trivial component from which the analog of equation \eqref{zeromy} follows.}. 
%%%%%
Such a friction term produces an energy lost in the zero mode characterized by equation \eqref{clavisima}. Therefore, our proposal is to identify this energy loss with the right hand side of \eqref{vlowy}, namely ${\rm stochastic \ source}\equiv \gamma H^5$ and write 
\be\label{mono}
{\dot W^{\rm pert.}}=\gamma H^5.\ee
%For convenience we rewrite \eqref{zeromy} here
%\be\label{lalan}
%\ddot \phi_0+3 H_0 \dot\phi_0+\frac{d V(\phi_0)}{d \phi} 
%%+\Gamma_\phi \dot\phi_0
%+\Gamma_{\rm Planck}\dot\phi_0 =0.
%\ee
This completes the conservation of energy picture in our scenario and we can write the full version of the continuity equations involving background as well as perturbations up to second order in perturbations. Concretely, using the definition \eqref{continuity}, and including the radiation generated by the decaying of the cosmological constant and the Higgs decay in other particles of the standard model, the continuity equation \eqref{inject}  becomes\ba\label{conditory}
&& \!\!\!\!\!\!\!\! \frac{d\left(\rho_\Lambda+\rho_{\rad}+\rho_{\rm h}+\sbraket{\delta\rho^{(2)}}\right)}{d\tau}+3H \left(\rho_\Lambda+\rho_{\rad}+\rho_{\rm h}+\sbraket{\delta\rho^{(2)}}+P_\Lambda+P_{\rad}+P_{\rm h}+\sbraket{\delta P^{(2)}}\right)=0\n \\
&& \!\!\!\!\!\!\!\! \frac{\dot \Lambda m_p^2}{8\pi}+\left(\dot \rho_{\rm rad}+4H \rho_{\rm rad}\right)+{\dot W^{\rm pert.}} +\dot \phi_0\left({\ddot \phi_0}+V^\prime(\phi_0)+3H \dot \phi_0\right)=0\n \\ &&\!\!\!\!\!\!\!\!
\underbrace{\frac{\dot \Lambda m_p^2}{8\pi}+\left(\dot \rho_{\rm rad}+4H \rho_{\rm rad}\right)
%-\Gamma_\phi \dot \phi^2_0
}_{=0}+\overbrace{{\dot W^{\rm pert.}}-\gamma  H^5}^{=0}=0,
\ea
where in going from the second to the last line we used the Higgs background equation \eqref{zeromy},  and we rearranged  the terms corresponding to the continuity equation for the perturbations replacing in addition \eqref{clavisima}.
The idea encoded in the previous equation is that the relaxation of the cosmological constant heats up radiation (which in the initial De Sitter phase dilutes exponentially and hence has negligible effect on the background dynamics) while the Brownian stochastic interaction of the Higgs rolling down the potential  produces fluctuations according the balance equation \eqref{mono} which as discussed in Footnote \ref{eleven} is an analog of Einstein's detail balance condition. 

The coefficient of friction $\gamma$ will be determined later from an  Einstein-like detailed balance condition that links the dissipation encoded in $\gamma$ with the amplitude of the observed power spectrum of fluctuations observed the CMB \footnote{Before calculating the power spectrum generated from the `detailed balance' equation \eqref{mono} we would like to comment on the fact that 
first order perturbations do not contribute to the ensemble average that led to our continuity equation \eqref{conditory}. What we have used (as first stated in \eqref{stochy}) is that, first order contributions to $T_{\mu\nu}$---while non-vanishing in a particular realization of the stochastic process (representing the particular state of our universe)---average to zero when considering an ensemble of realizations (ensemble of universes). But how can that be relevant for our own particular universe that is one among the members of the ensemble? The answer invokes an analogy with the ergodic hypothesis: the condition $\sbraket{\delta\phi(x)}=0$ is to be interpreted on a single realization (via this ergodicity assumption) as implying that, at a given time,  the space average 
\be
\frac{\int_{R} \delta\phi(\vec x,t) dx^3}{V_R}=0,
\ee
for a sufficiently large region $R$ (here $V_R$ is the co-moving volume of the region). In this way, the local contribution of fluctuations to $T_{\mu\nu}$ is not vanishing
in a given realization. Nevertheless, they average to zero in such the mean field sense. Similar interpretational questions arise for the ensemble average of the quadratic contributions to $T_{\mu\nu}$ when translated to our (single realization) universe. However, these are familiar issues common to conventional situations (see for instance \cite{Peter:1208401}).
%Section 5.1.3.3 in Peter (cosmic variance)
}. To leading order, such steady injection of energy in the inhomogeneities is at the heart of the scale invariant nature of 
power spectrum of density perturbations produced by this means. This is what we do in the next section.

\subsubsection{The power spectrum from diffusion}
Equation \eqref{keyly} shows that the stochastic ensemble expectation value of the product of scalar field fluctuations at a single point is directly related to the 
power spectrum of the fluctuations. This provides a simple relation between the expectation value of the energy momentum tensor and the power spectrum of the scalar field perturbations.
For instance, using \eqref{pdm}, an algebraic manipulation analogous to the one leading to \eqref{ps} implies 
\ba\label{keyl}
\sbraket{\delta \rho^{(2)}}\equiv \sbraket{\delta T_{00}} &=&\frac1{2\pi^2}\int dk k^2 \left(\frac{1}{2}P_{\delta\dot \phi}+\left[ \frac{k^2}{2a^2} + \frac{1}{2} \frac{d^2 V(\phi_0)}{d \phi^2}  \right]P_{\delta\phi}\right) \n \\
 &\approx &\frac1{2\pi^2}\int dk k^2 \left(\frac{k^2}{2 a^2}+ \frac{1}{2} \frac{d^2 V(\phi_0)}{d \phi^2}  \right)P_{\delta\phi},
\ea
where $P_{\delta\dot \phi}$ is defined via the analog of equation \eqref{ps} but for the fluctuations $\sbraket{\delta \dot \phi_{\vec k}\delta \dot\phi_{\vec q}}$, we assume that the stochastic process is isotropic (so that $dk^3\to 4\pi k^2 dk$), and  we neglected $P_{\delta\dot \phi}=\sO(a^{-6})$ due to \eqref{nobep}.
%,  and we expressed the contribution from the Higgs potential to leading order in $\lambda$ using \eqref{higgy} and the parameter $\epsilon_\phi=\phi_0/H_0$ introduced in \eqref{ef}. 
According to our previous discussion, we assume that the perturbations are created when the modes have physical wavelengths of the order of the Planck scale. In terms of the wave number this happens when $k \sim a m_p$ and we assume that modes are simply cut-off for large wave numbers, namely 
$P_{\delta\phi}(k)=0$ for $k>a m_p$. Thus, including this input in the integration boundaries of \eqref{keyl} we obtain
\ba\label{54}
\sbraket{\delta \rho^{(2)}}\equiv \sbraket{\delta T_{00}}&\approx&\frac1{2\pi^2}\int_{\mu}^{a m_p} dk k^2  \left(\frac{k^2}{2 a^2}+\frac{1}{2} \frac{d^2 V(\phi_0)}{d \phi^2} \right)P_{\delta\phi}(k),
\ea
where $\mu$ is an infrared cut-off that will not have any effect in the equations describing the regime of interest. Changing time variables from $\tau$ to $a(\tau)$, equation \eqref{mono} becomes 
\ba\label{condi}
\frac{dW^{\rm pert.}}{da}\equiv \frac{d \sbraket{\delta \rho^{(2)}}}{da}+\frac{2}{a} \sbraket{\delta \rho^{(2)}} -\frac{1}{a}\frac{d^2 V(\phi_0)}{d \phi^2}  \sbraket{\delta\phi^2 }
%&=&H_0 \left(\left. \frac{k^2}{4\pi^2} \left(\frac{k^2}{a^2}+\lambda H_0^2\epsilon_\phi^2 \right) P_{\delta\phi}\right|_{k=aH_0}\right)-\frac 2 {4\pi^2 a}\int_0^{aH_0} dk k^2  \frac{k^2}{a^2}P_{\delta\phi}\n \\
&=&\gamma \frac{H^4}{a}.\ea
To leading order in $\lambda$, the previous equation tell us that the amount of energy that we are extracting from the Higgs zero mode to produce inhomogeneities is done in a way that is not sensitive to the size of the universe. More precisely $dW^{\rm pert.}= W_0 (da/a)$ with $W_0=\gamma  H^4$ is a self-similar process (invariant under rescaling $a\to \alpha a$) to leading order in $\lambda$ (recall \eqref{higgy}). This is of course consistent with the assumption that led, via dimensional analysis, to equation \eqref{clavisima}.  

Using equation \eqref{54} and \eqref{higgy} one can substitute the ansatz $P_{\delta\phi}(k)=P_0/k^3(1 +\sO(\lambda))$ into \eqref{condi} and check that it produces a solution of the detailed balance condition to leading order in $\lambda$.  Thus, in the present model the Harrison-Zeldovich spectrum of inhomogeneities in the scalar field can be simply related to a self-similar injection of energy during the quasi-inflationary era $H \approx $ constant without the need to invoke the uncertainty principle and (most importantly) the pre-existence of vacuum fluctuations as described by the extrapolation of quantum field theory to trans-Planckian scales. The solution is 
\be\label{solution}
P_{\delta\phi} (k)=\frac{P_0}{k^3} ,
\ee
with  \be\label{cafe}
P_0={4}\pi^2 \gamma   \frac{H_0^3}{m_p} \left(1-6\lambda\right)\approx {4}\pi^2 \gamma \frac{H_0^3}{m_p},
\ee 
where the next to leading order correction is not relevant when comparing with observations because it does not depend on $k$; hence, we drop it for simplicity.
However, we will see in the following section that the $\lambda$ corrections will affect scale invariance when one instead analyses the effects of these perturbations in the gravitational field (which are the ones directly related to the observed fluctuations in the CMB). In fact, the red tilt of the CMB power spectrum is linked (in this model) to the self-interaction strength of the Higgs field $\lambda$.

%%%%%%%%%%%%%%%%%%%%%%%%%%%%%%%%%%%%%%%%%%%
%IN DERIVING THE PREVIOUS RESULT WE USE PERTURBATION THEORY AND GO ONLY UP TO FIRST ORDER, AT THAT ORDER THE "BULK" TERM (THE ONE THAT IS GIVEN BY AN INTEGRAL IN K) IS IDENTICALLY ZERO IN THE BALANCE EQUATION THAT RELATED THE FRICTION IN THE ZERO MODE OF THE HIGGS WITH THE FLUCTUATIONS. THIS SIMPLIFIES THE ANALYSIS OF THE MECHANISM. HIGHER ORDER CORRECTIONS COULD BE CALCULATED IN  PRINCIPLE BUT MORE ANALYSIS WOULD BE NEEDED.
%%%%%%%%%%%%%%%%%%%%%%%%%%%%%%%%%%%%%%%%%%%

\subsubsection{The Weinberg theorem and the power spectrum of density fluctuations at the CMB}\label{resultados}

The following equations concern long wavelength modes $k<aH_0$ (those added up in \eqref{54}). 
Weinberg proved a beautiful and very powerful statement concerning such modes based on the universality of free-fall. This result is know as Weinberg's theorem \cite{Weinberg:2008zzc};  the proof of which is revisited and simplified in the Appendix \ref{WT}. One has in particular that the gravitational potential for these super Hubble modes is given by (see \eqref{huy1})
\be\label{desitete}
\Phi_k=\Psi_k=\mathcal{R}_k\left(-1+\frac{H(\tau)}{a(\tau) } \int_{\mathcal{T}}^{\tau} a\left(\tau^{\prime}\right) d \tau^{\prime}\right)\approx 0,
\ee
where $\mathcal{R}_k$ are time independent, and the right hand side approximation is valid during the De Sitter phase.
This implies that scalar perturbations do not generate scalar metric perturbations during the inflationary era (they decouple gravitationally to leading-order-perturbation-theory in $\lambda$ when one has an almost exact De Sitter phase).
This justifies the omission (for presentation simplicity) of the scalar metric perturbations in the expression of the linear expansion to the energy momentum tensor \eqref{ppp}.
The form of the scalar density fluctuations including the gravitational potential are given below in \eqref{posta}.

Weinberg's theorem also implies the existence of adiabatic scalar density perturbations as solutions of linearized gravity for which
\begin{equation}
	\frac{\delta \rho^{(1)}_k}{{{\dot\rho_0}}}=-\frac{\mathcal{R}_k}{a(\tau)} \int_{\mathcal{T}}^{\tau} a\left(\tau^{\prime}\right) d \tau^{\prime} \approx -\frac{\mathcal{R}_k}{H_0},
\end{equation}
where these are adiabatic in that the previous relation for each species contributing to the perturbations individually (see \eqref{huy3}).   
The previous two equations correspond to  equations (5.4.4) and (5.4.5) in \cite{Weinberg:2008zzc} and will be discussed and recovered in Appendix \ref{WT}. In the present context they allow us to compute $\mathcal{R}_k$ for super-Hubble scales $k<aH_0$ as
\be\label{eity1}
\mathcal{R}_k=-H_0 \frac{\delta \rho^{(1)}_k}{{{\dot\rho_0}}}.
\ee
We have that (recall \eqref{pipi})
\be
\rho_0=\frac{\Lambda m_p^2}{8\pi}+\frac{\dot \phi^2_0}2+V(\phi_0)+\rho_{\rm rad}.
\ee
 Assuming that $\Lambda\approx$ constant during the De Sitter phase, and using the field equations \eqref{zeromy} and the equation of state of the radiation component,  the time derivative of $\rho_0$ gives
\ba\label{pista}
\dot \rho_0 &=&{\dot \phi_0} \ddot\phi_0+\frac{dV(\phi_0)}{d\phi_0} \dot\phi_0-4 H_0 \rho_{\rm rad}\n \\
&=&-{\dot \phi_0} (3 H_0 \dot\phi_0+\frac{d V(\phi_0)}{d \phi} +\Gamma_{\rm Planck}\dot\phi_0 )+\frac{dV(\phi_0)}{d\phi_0} \dot\phi_0-4 H_0 \rho_{\rm rad} \n \\
&\approx&-H_0 \left[3 \dot\phi^2_0 + 4 \rho_{\rm rad}\right]\n \\
&\approx&
%-H_0 \left[ \frac{12}{9}\left(1+\frac{\alpha^\star}{3}\right){ \lambda^2 \epsilon_\phi^6} {H_0^4} + 4 \rho_{\rm rad}\right]\approx
 - \frac{4}{3H_0}{\lambda^2 } {\phi_0^6},
\ea
where in the last line we use that $\Gamma_{\rm Planck}\ll H_0$ (to be confirmed below), and that $\rho_{\rm rad}\ll \lambda {\phi_0^4}$ (this requirement can be met if one is in the initial range of of De Sitter evolution where radiation is exponentially diluted, see Figure \ref{fig:loglog}).
%, and it is also necessary that we ignore such degrees of freedom so that the correct counting of degrees of freedom leads to adiabaticity from Weinberg's conservation law \cite{Weinberg:2008zzc}). 
From the general expression of the energy-momentum tensor we get (including the metric perturbation term) %LA EQUATION SIGUEINTE LA HICIMOS A MANO PERO TAMBIEN ESTA EN 10.1.9 de Weinberg
\ba\label{posta}
\delta\rho^{(1)}_k&=&\dot \phi_0 \delta\dot\phi_k(x^\mu)+V^\prime(\phi_0)\delta\phi_k(x^\mu)-\Psi_k \dot \phi^2_0\n \\
&=& V^\prime(\phi_0)\left(-\frac{\delta\dot\phi(x^\mu)}{3H_0}+\delta\phi(x^\mu)\right)-\Psi_k \dot \phi^2_0\n \\
&\approx& 2\lambda \phi_0^3 \delta\phi_k(x^\mu),
\ea
where we neglect the $\delta\dot\phi_k$ term as it quickly dies off for super-Hubble modes according to \eqref{death}, and we used that the long wavelength adiabatic scalar metric perturbations vanish in the De Sitter phase according to \eqref{desitete}.
Replacing \eqref{posta} and \eqref{pista} in \eqref{eity1} we get
\ba\label{fgfg}
\mathcal R_k&=& \frac{3 H_0^2}{{2}{  } {\lambda  \phi^3_0}} \delta\phi_k\n \\
&=&\frac{3}{{2}{ } {\lambda \epsilon^3_\phi H_0 }}  \left[1+{2}\lambda \log\left(\frac k {H_0}\right)\right]{\delta\phi_k},
\ea
where we have used the expression on the right of \eqref{higgy} and used that the modes k are generated at horizon crossing when $a=k/H_0$.
Squaring the previous relationship and computing its ensemble average in our stochastic process one obtains, from the definition \eqref{ps}, an equation linking the power spectrum $P_{\mathcal R}$ of  the ${\mathcal R}_k$ and that of the scalar perturbations. Explicitly, using \eqref{cafe}, we get 
\ba\label{poweryes}
P_{\mathcal R}&=& \frac{9 }{{4}{ \lambda^2 } {H_0^2  }}\frac{P_0}{k^3}  \left[1+{{4}\lambda \log\left(\frac k{k_0}\right)-4 \lambda \log\left(\frac {H_0}{k_0} \right) }\right]\n \\
%&=& \frac{9 }{{4}{ \lambda^2  \epsilon_\phi^6} {H_0^2}} \frac{{16}\pi^2 \gamma \lambda^2  \epsilon_\phi^6 H_0^2 \left(1-6\lambda \epsilon_\phi^2\right)}{9k^3} \left(1+4\lambda\epsilon^2_\phi \log\left(\frac k{k_0}\right)-4\lambda\epsilon^2_\phi \log\left(\frac {H_0}{k_0} \right)\right)\n \\
&=& \frac{9 \pi^2 \gamma}{k^3 \lambda^2 } \left[1+{{4}\lambda \log\left(\frac k{k_0}\right)-4 \lambda \log\left(\frac {H_0}{k_0} \right) }\right]
.\ea
If we take $H_0/k_0=1$ which boils down to normalizing $a=1$ at the moment the most IR mode in the CMB leaves the horizon we arrive at the final expression for the power spectrum of scalar perturbations (for $H_0\approx m_p$) we get 
\be\label{boxed}
{P_{\mathcal R}\approx  \frac{9 \pi^2 \gamma }{k^3\lambda^2}\left(1+ 4 \lambda \log\left(\frac k{k_0}\right)\right).}
\ee
Using the customary notation where $P_{\mathcal R}\equiv N^2/k^3$,  comparison with  CMB observations (see for instance \cite{Weinberg:2008zzc}) fixes the normalization factor $N^2$   to
\be\label{gamita}
N^2\approx \frac{9 \pi^2 \gamma}{\lambda^2} \approx 1.9 \times 10^{-10}.
\ee	
Using that $\lambda\approx -10^{-2}$ at our energy scale one needs to fix $\gamma\approx 10^{-16}$ which is remarkably close to the
estimate $\gamma_{\rm H}$ given in \eqref{conficinfi} based on the natural measure of deviation from conformal invariance 
put forward in the introduction expected to control the Brownian diffusion mechanism. Deviation from scale invariance are encoded in the spectral index of scalar perturbations $n_{\rm s}$. They are controlled by the Higgs self coupling as it follows from 
\eqref{boxed}. The result to first order in $\lambda$ is 
\be\label{indexe}
n_{\rm s}-1\equiv \frac{d\log(k^3 P_{\mathcal R})}{d\log k}\approx 4\lambda +\sO\left[\lambda^2 \log\left(\frac{k_{\rm max}}{k_0}\right)\right].
\ee
Observations constraint it to \be 1-n_{\rm s}=0.04\pm 0.004,\ee which implies $\lambda\approx -\, 10^{-2}$ which is compatible with the he standard model expected value of $\lambda=-(1.3\pm 0.7)\times 10^{-2}$ at these high field values---see \cite{Degrassi:2012ry}.  
Notice that in our framework the spectral index is itself $k$ dependent. Notice that the linear approximation used remains consistent inspite of the $ \log({k_{\rm max}}/{k_0})$ in the error term as for  $\lambda=-10^{-2}$ and $k_{\rm max} =10^5 k_0$  one has $\lambda^2\log({k_{\rm max}}/{k_0})\approx 10^{-3}$ which is smaller than the present observational error in $1-n_{\rm s}$ \cite{Akrami:2018odb}. In the same paper the deviations from a constant spectral index are reported to be given by 
\be
\frac{dn_{\rm s}}{d\log k}=-0.0045\pm 0.0067.
\ee 
One can repeat the previous analysis starting from equation \eqref{54} and keeping terms up to order $\lambda^2$. 
With this improved approximation it is possible to compute the previous quantity and the result is
\be
\frac{dn_{\rm s}}{d\log k}=-0.0005+\sO(\lambda^3).
\ee 
The previous is a prediction of our scheme, potentially verifiable in the future if observational data 
reduce the error by about $10\%$. 

Finally, notice that only the first and second moments of the probability distribution of our stochastic process have entered our analysis (equations (\ref{stochy}) and \eqref{keyly}). It would be interesting to study the possible implications of higher non-trivial moments (for instance starting with the assumption of non vanishing third moments) and the associated deviations from Gaussianity. This is out of the scope of the present work but probably worth of more dedicated consideration.

\subsubsection*{Tensor modes}

So far we  focused on the description of a mechanism for the generation of inhomogeneities in scalar modes only.
The question of whether tensor modes are also produced is a very important one in view of future constraints on the 
scalar-to-tensor ratio $r$ from CMB observations. In our model fundamental discreteness is the underlying mechanism for the active generation of the inhomogeneities.  As argued in the introduction, see also  \cite{Perez:2017krv, Perez:2018wlo} for further discussion,  such discreteness should primarily affect degrees of freedom 
 breaking scale invariance. In the present case, with the assumption of the validity of the standard model, the breaking of scale invariance is mediated by the Higgs scalar mass. Gravitons being massless should not interact with the Planckian discrete structure according to the dimensional analysis arguments behind the construction of our model. More precisely, as it is well known, an infinitesimal conformal transformation $\delta g_{ab}= \delta \omega g_{ab}$---here regarded as a field variation---leads to the trace-part of Einsteins equations $(R-8\pi G T)=0$. This relates the trace part of the Einstein field equations with the conformal-invariant-breaking interactions, which are those that mediate the stochastic production of inhomogeneities in our model.  Thus the Planckian granularity---imposed by the consistency with the low energy Lorentz invariance \cite{PhysRevLett.93.191301, Perez:2018wlo}---cannot generate tensor modes whose sources are encoded in the tensor traceless components of the energy momentum tensor. Therefore, the expected value of the tensor-to-scalar ratio predicted by our model to be highly supressed, i.e.,  $r\approx 0$. It would certainly be interesting to have a more quantitative estimate of $r$, but this might require better understanding of the quantum geometry dynamics close to the fundamental scale as the effect sought is a sub leading one (gravitons being massless in the low energy description).  We leave this for future investigations.

\section{Planckian black hole remnants as dark matter}\label{de}

The fundamental nature of dark matter remains and open question. Here we would like to stress that, if the reheating temperature at the end of inflation gets close to the Planck temperature, a model of dark matter where it is made of quantum gravity Planck mass particles (as described from our low energy perspective) that only interact gravitationally is very natural. 

Little is known about the fundamental theory of quantum gravity besides the fact that it has to reproduce general relativity with massless gravitons a low energies. As emphasized in the introduction  several approaches to quantum gravity propose that the smooth geometry of general relativity would be emergent from an underlying fundamental discrete structure at the Planck scale. In these approaches the fundamental energy scale $m_p$ plays a central role. A point we would like to stress here is that, in addition to motivating the mechanism for generation of structure studied in this paper, such perspective naturally leads to the possibility that defect-like objects in the discrete fabric spacetime could survive the continuum limit. If so it seems likely that these would behave like particles with a mass with the natural mass scale $m_p$ and would  interact  only gravitationally.  Such defects could be thermally excited if Planckian temperatures were achieved during reheating. It is unclear how to picture such particles  from our low energy perspective, for the lack of a better name we could think of them as Planckian stable primordial black holes. Unstability of tiny black holes due to Hawking radiation is often evoked to rule out such dark matter candidates. However, lacking a full quantum gravity theory, it is clear that little is certain about the properties of black holes (or such Planckian defects) of that scale. It is even unclear in what sense such objects qualify as black holes when the very notion of geometry is expected no to be available so close to the fundamental scale.  The only thing that is certain in fact is that absolutely all the assumptions behind Hawking's calculation simply fail: thus the simple invocation of Hawking radiation is not a serious argument to rule out their hypothetical role in cosmology.  

The possibility that dark matter is made of primordial Planckian black hole remnants (or more humbly, Planck mass purely gravitationally interacting particles) has been evoked in the literature before \cite{pmpbh, Chen:2002tu, Rasanen:2018fom, Rovelli:2018okm}. Here we show that the dark matter energy density required by observations can indeed emerge naturally in a Planck scale reheating scenario as the one produced by our model. Such type of dark matter will basically behave like a dust fluid interacting with the rest of matter gravitationally only. It would be extremely hard to detect via other manifestations. Their presence would remain hard to notice locally as the Planckian size of these particles will make their gravitational cross section in interactions with usual matter extremely small (however,  this form of dark matter might be directly detectable via its gravitational interaction \cite{Carney:2019pza}).

At the end of the inflationary era reheating raises the temperature to close to the Planck temperature and Planck mass remnants could be created via thermal fluctuations if thermal equilibrium density is achieved. In order for this to happen one needs the remnant interaction rate $\Gamma_{\rm pbh}>H$, where the interaction rate is given by $\Gamma=n \sigma v$ with $n$ the number density, $\sigma$ the interaction cross section, and $v$ the velocity.  For  remnants of mass $m_{\rm pbh}$ the interaction cross section is $\sigma_{\rm pbh}\approx m_{\rm pbh}^2/m_p^4$  while their number density $n$  while in thermal equilibrium goes like $n\approx T^3$. Using that in the radiation dominated era $H\approx (T/m_p) T$,  we conclude that remnants decouple from thermal equilibrium when 
\ba
% No need to care about prefactors in this analysis, we are deep into the QG regime
%\n H\approx (T/m_p) T> T^3  m^2/m_p^4\\
%\n H\approx 1 > T  m^2/m_p^3\\
 T \lesssim  \frac{m^2_p}{m_{\rm pbh}^2}\, m_p\equiv T_{\rm D}.
\ea
If thermal equilibrium can hold up to $T_{\rm D}\lesssim T_{\rm end}$ then the thermal  remnant
abundance of dark matter today can be estimated to be about (see equation 4.38 in \cite{Peter:1208401})
%Equation 4.38 Peter
\be
\frac{\rho^{\rm thermal}_{\rm pbh} (T_{\rm D})}{m_p^4}\approx  \left(\frac{m_{\rm pbh}}{m_p}\right)^4\left( \frac{T_{\rm today}}{T_{\rm D}}\right)^3 \left(\frac{T_{\rm D}}{m_{\rm pbh}}\right)^{\frac{3}{2}}  e^{-\frac{m_{\rm pbh}}{T_{\rm D}}}.
\ee 
%From decoupling until today the density of dark matter particles decays  by a factor of $a^3_{\rm D}/a^3_{\rm today}\approx T^3_{\rm today}/T^3_{\rm D}=$
%% ESTIMATE MAS EXPLICITO
%\be
%\frac{\rho^{\rm thermal}_{\rm pbh} ({\rm today})}{m_p^4}\approx  \left(\frac{m_{p}}{m_{\rm pbh}}\right)^{\frac{1}{2}}  e^{-\left(\frac{m_{\rm pbh}}{m_p}\right)^3}=10^{-20}.
%\ee 
One can easily check that it is possible to obtain a remnant density compatible  with dark energy density today---which would correspond to evaluating the previous line to about $10^{-120}$---with a $m_{\rm pbh}$ slightly larger than but of the order of $m_p$. This shows that the framework provided by our model could also fit dark energy genesis from the production of stable PBHs via thermal fluctuations at the end of the De Sitter phase without extreme fine tuning where the necessary suppression is brought by the standard Gibbs factor. After completion of this work we discovered that very similar arguments are put forward in \cite{Barrau:2019cuo}.

\section{Discussion} \label{dis}

We have proposed a model where the cosmological constant  $\Lambda_0$ starts off with its natural Planckian value and later relaxes via diffusion into the matter degrees of freedom while driving an inflationary era. We assumed that the cosmological constant decays exponentially in unimodular time which leads to the necessary number of e-folds if the parameter $\beta$ is sufficiently small. However, all the observational predictions of the model are independent of the precise value of $\beta$ as long as it is sufficiently small. The validity of our analysis requires only that the cosmological constant remains Planckian for a minimum number of e-folds (Section \ref{efo}). The standard model of particles physics is assumed to be valid all the way to close to the Planck scale and the Higgs scalar also assumed to start with a large semiclassical value $\phi_0$ close to the Planck scale. The initial conditions of the other matter components do not affect the dynamics in any important manner as long as the radiation density is not ultra-Planckian (as in standard inflation \cite{Martin:2013tda}, the cosmological constant dominates and the expansion dilutes away any memory of these initial conditions). The relaxation mechanism is associated with the hypothesis of discreteness of quantum gravity at the Planck scale. This suggest a natural time variable proportional to the number of Planckian four volume elements created by the dynamical evolution and in terms of which the relaxation is exponential. 
We argue that the same underlying discreteness at about the Hubble scale $H_0$ should stimulate the generation of inhomogeneities in the Higgs amplitude at that very scale, and show that a stochastic model where the steady injection of energy at the Hubble scale 
produces (to leading order in the Higgs self coupling $\lambda$) a scale invariant spectrum of density perturbations with an amplitude that is compatible in order of magnitude with CMB observations. 

More precisely, once the initial values of the Higgs background and the cosmological constant are fixed to the natural scale $m_p$ the model is controlled by two parameters: the parameter $\beta$ which defines the decay rate of the cosmological constant in unimodular time, and the parameter $\gamma$ parametrizing the Ohmian friction term---stemming from the interaction with discreteness exciting inhomogeneities---in the field equations for the zero model of the Higgs. As mentioned above, the parameter $\beta$ needs only to be sufficiently small in order to achieve a sufficient number of e-folds that makes the model compatible with observations (fixing $\beta$ amount to fixing the number of e-folds of inflation).  The parameter $\gamma$ is a dimesionless coupling representing noisy interaction of the Higgs with the granular structure at the Planck scale which in turn is expected to be possible thanks to the breaking of scale invariance of the Higgs scalar. The natural order parameter for such breaking is $\gamma_{\rm H}\equiv m_{\rm H}/m_p$. It is a remarkable fact that agreement with the observation of the perturbations at the CMB   necessitates a $\gamma\approx 10^{-16}$ which coincides (in order of magnitude) with $\gamma_{\rm H}$.

Deviations from scale invariance are brought by the evolution of the Higgs on the Higgs potential and depend on $\lambda$. Remarkably, standard model physics (encoded in $\Lambda$) produces a red tilt of the spectrum that is in agreement with the data extracted from the CMB observations: the spectral $n_{\rm S}$ coincides with observations for $\lambda\approx -10^{-2}$ which is compatible with the expected value of $\lambda$ at high energies in the standard model. Moreover, the model predicts a variation of the spectral index with scale 
that is inside the limits obtained from the analysis of latest data \cite{Akrami:2018odb}. This corresponds to a proper prediction of our analysis which 
could be tested in the future if observational errors are reduced by an order of magnitude.

Given the above mentioned initial conditions, Planckian temperature reheating is a robust ($\beta$-independent) prediction of our model.  We observe that 
such a feature  could naturally account for the present abundance of dark matter via the thermal production of Planck mass defects if such stable particles are part of the spectrum of quantum gravity. As in the case of the so-called WIMP miracle, we notice that the decoupling temperature and mass of such hypothetical purely gravitationally interacting particles (natural objects from the perspective of quantum gravity) fall in the right range to represent a possible dark matter candidate. 

We are aware of the strong assumptions in our model which stretches well established physics into the uncertain and unknown territory of quantum gravity. The speculative nature of such an enterprise is certainly very risky. Our model links naturally some of the key cosmological observations with aspects of that new physics of quantum gravity that we strive to better understand. This by itself seems to justify our adventures. We hope that these initial ideas could lead to helpful insights in the future. 
 
%  Warm inflation \cite{PhysRevLett.75.3218}.  

\section*{Aknowledgements}

We are grateful for exchanges with Daniel Boyanowsky,  Chris Korthals Altes, Marc Knecht, and Federico Piazza at different stages of this work. 
We thank Robert Brandenberger for remarks that  helped us improve the presentation of the similarities and differences between our model and the standard account of the mechanism for  the generation of inhomogeneities during inflation. We acknowledge the work of one anonymous referee for the meticulous reading of our manuscript, the questions raised, and the suggestions made which have helped improving the quality of our presentation.  
The present work has slowly emerged from long discussions with Daniel Sudarsky, many of the key features of our proposal come from this interaction and from the effort in circumventing some of his criticisms. A.P. would like to acknowledge the immense hospitality of the people of the {\em Vall\'ee Fran\c{c}aise} where some of the key ideas leading to this work were found.

\begin{appendix}

\section{Revisiting the Weinberg theorem} \label{WT}

This is a short review of some basic facts of cosmological perturbation theory and the proof of Weinberg conservation theorem. We follow the notation of \cite{Weinberg:2008zzc}. The proof presented here is, we believe, more direct than the one in the textbook; we include it here for completeness. In perturbation theory, the metric is split in the usual way as
\begin{equation}\label{eqbox1}
g_{\mu \nu} = {g}^{(0)}_{\mu \nu} + h_{\mu \nu}
\end{equation}
where ${g}^{(0)}_{\mu \nu} $ is the unperturbed, $K=0$ metric, and $h_{\mu \nu}$ is a perturbation; namely 
\begin{equation}\label{eqbox1.2}
{ds}_0^{2} = - d\tau^{2} + a(\tau)^{2} \delta_{i j}dx^{i} dx^{j}.
\end{equation}
The metric perturbation $h_{\mu \nu}$ can be decomposed as
\begin{equation}\label{eqbox2}
\begin{aligned}
h_{00} &=-E \\
h_{i 0} &=a\left[\frac{\partial F}{\partial x^{i}}+G_{i}\right] \\
h_{i j} &=a^{2}\left[A \delta_{i j}+\frac{\partial^{2} B}{\partial x^{i} \partial x^{j}}+\frac{\partial C_{i}}{\partial x^{j}}+\frac{\partial C_{j}}{\partial x^{i}}+D_{i j}\right],
\end{aligned}
\end{equation}
where $(A,B,E,F)$, $(G_{i},C_{i})$   and $D_{i j}$ are scalar, vector and tensor degrees of freedom respectively, satisfying the conditions 
\begin{equation}\label{eqbox3}
\frac{\partial C_{i}}{\partial x^{i}}=\frac{\partial G_{i}}{\partial x^{i}}=0, \quad \frac{\partial D_{i j}}{\partial x^{i}}=0, \quad D_{i i}=0.
\end{equation}
In the same way we consider first order perturbations to the energy momentum tensor
\begin{equation}\label{eqbox4}
\begin{aligned}
\delta T_{i j}&=p h_{i j}+a^{2}\left[\delta_{i j} \delta p+\partial_{i} \partial_{j} \pi^{S}+\partial_{i} \pi_{j}^{V}+\partial_{j} \pi_{i}^{V}+\pi_{i j}^{T}\right] \\
\delta T_{i 0}&=p h_{i 0}-(\rho+p)\left(\partial_{i} \delta u+\delta u_{i}^{V}\right) \\
\delta T_{00}&=-\rho h_{00}+\delta \rho,
\end{aligned}
\end{equation}
where $\rho$ and $p$ denote the zero order values (the background values), and $\delta \rho$, $\delta p$, $\delta u$, $\delta u_i^{V}$,   $\pi_{i}^{V}$, and $ \pi_{i j}^{T}$ are perturbations of the density, the pressure, the fluid flow vector field, and the stresses respectively. They satisfy the usual conditions 
\begin{equation}\label{eqbox5}
\partial_{i} \pi_{i}^{V}=\partial_{i} \delta u_{i}^{V}=0, \quad \partial_{i} \pi_{i j}^{T}=0, \quad \pi_{i i}^{T}=0.
\end{equation}
Due to the symmetries of the background it is possible to write the linearised field equations as a set of decoupled equations for scalar, vector and tensor modes. For the argument leading to Weinberg's conservation laws we will only need the equations for scalar and tensor modes.

\vskip.2cm 
\paragraph{Scalar Modes:}
The equations governing the scalar modes are
\begin{equation}\label{eqbox6}
\begin{aligned}
-4 \pi G a^{2}\left[\delta \rho-\delta p-\nabla^{2} \pi^{S}\right]=& \frac{1}{2} a \dot{a} \dot{E}+\left(2 \dot{a}^{2}+a \ddot{a}\right) E+\frac{1}{2} \nabla^{2} A-\frac{1}{2} a^{2} \ddot{A} \\
&-3 a \dot{a} \dot{A}-\frac{1}{2} a \dot{a} \nabla^{2} \dot{B}+\dot{a} \nabla^{2} F, \\
\end{aligned}
\end{equation}
\begin{equation}\label{eqbox7}
\partial_{j} \partial_{k}\left[16 \pi G a^{2} \pi^{S}+E+A-a^{2} \ddot{B}-3 a \dot{a} \dot{B}+2 a \dot{F}+4 \dot{a} F\right]=0,
\end{equation}
\begin{equation}\label{eqbox8}
8 \pi G a(\rho+p) \partial_{j} \delta u=-\dot{a} \partial_{j} E+a \partial_{j} \dot{A},
\end{equation}
\begin{equation}\label{eqbox9}
\begin{aligned}
-4 \pi G\left(\delta \rho+3 \delta p+\nabla^{2} \pi^{S}\right)=-\frac{1}{2 a^{2}} \nabla^{2} E-\frac{3 \dot{a}}{2 a} \dot{E}-\frac{1}{a} \nabla^{2} \dot{F}-\frac{\dot{a}}{a^{2}} \nabla^{2} F \\
+\frac{3}{2} \ddot{A}+\frac{3 \dot{a}}{a} \dot{A}-\frac{3 \ddot{a}}{a} E+\frac{1}{2} \nabla^{2} \ddot{B}+\frac{\dot{a}}{a} \nabla^{2} \dot{B},
\end{aligned}
\end{equation}
with the energy-momentum conservation equations 
\begin{equation}\label{eqbox10}
\partial_{j}\left[\delta p+\nabla^{2} \pi^{S}+\partial_{0}[(\rho+p) \delta u]+\frac{3 \dot{a}}{a}(\rho+p) \delta u+\frac{1}{2}(\rho+p) E\right]=0,
\end{equation}
\begin{equation}\label{eqbox11}
\begin{aligned}
\delta \dot{\rho}+\frac{3 \dot{a}}{a}(\delta \rho+\delta p)+\nabla^{2}\left[-a^{-1}(\rho+p) F+a^{-2}(\rho+p) \delta u+\frac{\dot{a}}{a} \pi^{S}\right] \\
\quad+\frac{1}{2}(\rho+p) \partial_{0}\left[3 A+\nabla^{2} B\right]=0.
\end{aligned}
\end{equation}

\paragraph{Tensor Modes:} For tensor modes we have only one equation
\begin{equation}\label{eqbox12}
-16 \pi G a^{2} \pi_{i j}^{T}=\nabla^{2} D_{i j}-a^{2} \ddot{D}_{i j}-3 a \dot{a} \dot{D}_{i j}.
\end{equation}

\subsection{Gauge transformations}
 A gauge transformation generated by an arbitrary vector field $\varepsilon^{\mu}(x)$ 
\begin{equation}\label{eqbox13}
x^{\mu} \rightarrow x^{\mu} + \varepsilon^{\mu}(x),
\end{equation}
 induces a transformation on the metric perturbation given by 
 \begin{equation}\label{eqbox14}
\begin{aligned}
\Delta h_{i j} &=-\frac{\partial \varepsilon_{i}}{\partial x^{j}}-\frac{\partial \varepsilon_{j}}{\partial x^{i}}+2 a \dot{a} \delta_{i j} \varepsilon_{0}, \\
\Delta h_{i 0}&=-\dot \varepsilon_{i} -\frac{\partial \varepsilon_{0}}{\partial x^{i}}+2 \frac{\dot{a}}{a} \varepsilon_{i}, \\
\Delta h_{00}&=-2 \dot \varepsilon_{0} .
\end{aligned}
\end{equation}
One can decompose the spatial part of the vector field $\varepsilon^{\mu}$ into a scalar part $\varepsilon^{S}$ and a divergenceless vector $\varepsilon_{i}^{V}$ :
\begin{equation}\label{eqbox15}
\varepsilon_{i}=\partial_{i} \varepsilon^{S}+\varepsilon_{i}^{V}, \quad \partial_{i} \varepsilon_{i}^{V}=0.
\end{equation}
Then the quantities defined in \eqref{eqbox2}  transform as
\begin{equation}\label{eqbox16}
\begin{aligned}
\Delta A=\frac{2 \dot{a}}{a} \varepsilon_{0}, \quad& \Delta B=-\frac{2}{a^{2}} \varepsilon^{S}, \\
\Delta C_{i}=-\frac{1}{a^{2}} \varepsilon_{i}^{V}, \quad& \Delta D_{i j}=0, \quad \Delta E=2 \dot{\varepsilon}_{0}, \\
\Delta F=\frac{1}{a}\left(-\varepsilon_{0}-\dot{\varepsilon}^{S}+\frac{2 \dot{a}}{a} \varepsilon^{S}\right), &\quad \Delta G_{i}=\frac{1}{a}\left(-\dot{\varepsilon}_{i}^{V}+\frac{2 \dot{a}}{a} \varepsilon_{i}^{V}\right).
\end{aligned}
\end{equation}

\subsection{The theorem}

To begin, let us concentrate on the scalar mode equations only. The Newtonian gauge is defined by setting $F=B=0$. 
However, we will keep the $F$ contributions as we will actually move away from the condition $F=0$ in what follows. However, we will change $F$ in a way that alter the scalar equations as written in the Newtonian gauge in a mild way (this is the key of the proof). One renames fields according to
\begin{equation}\label{eqbox17}
E \equiv 2 \Phi, \quad A \equiv-2 \Psi.
\end{equation}
The scalar field equations in the Newtonian gauge become (when $\pi_s=0$)
\ba 
-4 \pi G a^{2} \left( \delta \rho-\delta p
%-\nabla^{2} \pi^{S}
\right)&=&a \dot{a} \dot{\Phi}+\left(4 \dot{a}^{2}+2 a \ddot{a}\right) \Phi-\nabla^{2} (\Psi-\dot{a} F)+a^{2} \ddot{\Psi} +6 a \dot{a} \dot{\Psi} \label{lapla}, \\
&& -8 %\pi G a^{2} \partial_{i} \partial_{j} \pi^{S}&=
\partial_{i} \partial_{j}[\Phi-\Psi+ a \dot{F}+2 \dot{a} F]=0  \label{ij},\\
4 \pi G a(\rho+p) \partial_{i} \delta u&=&-\dot{a} \partial_{i} \Phi-a \partial_{i} \dot{\Psi} \label{ident},\\
4 \pi G\left(\delta \rho+3 \delta p
%+\nabla^{2} \pi^{S}
\right)&=&\frac{1}{a^{2}} \nabla^{2} (\Phi +a \dot{F}+{\dot{a}} F )+\frac{3 \dot{a}}{a} \dot{\Phi} +3 \ddot{\Psi}+\frac{6 \dot{a}}{a} \dot{\Psi}+\frac{6 \ddot{a}}{a} \Phi 
\label{lalila}.\ea
We first do a gauge transformation $\epsilon^{\mu}=(\epsilon_0(x^\mu), 0,0,0)$ on the background geometry (i.e., we have $h_{\mu\nu}=0$ to begin with). This gauge transformation---a simple time reparametrization---maintains the Newtonian-gauge-condition $B=0$ while it breaks the other Newtonian-gauge-condition by sending  $F=0\to F=- \epsilon_0/a$. Thus, the gauge transformation yields the following values to the (pure gauge) scalar perturbations
\ba\label{gygy}
F&=&-\frac{ \epsilon_0}a\n, \\ 
\Phi&=&\dot \epsilon_0\n, \\
 \Psi&=& -\frac{\dot a}{a} \epsilon_0\n, \\
 B &=&0.
\ea
This implies that $\Phi +a \dot{F}+{\dot{a}} F=0$ and $\Psi-\dot{a} F=0$.
%Therefore, aside from the equation \eqref{ij} where $F$ remains,  the form of the other equations is unchanged from Newtonian gauge perturbation equations. 

The previous gauge transformation takes us away from the Newtonian gauge for which $F$ should remain equal to zero. 
However, the perturbation equations can be rewritten in the Newtonian gauge form  by means of an additional transformation which (this time) is not a gauge transformation. This is how from a pure gauge mode, defined by (\ref{gygy}), a new physical perturbation---solution of the linearized equations---emerges.
In order to do this we first demand that the combination (appearing in \eqref{ij})  \be\label{marita} a { \dot F}+2 \dot{a}  F=\mathcal{R}\ee for some constant $\mathcal{R}$.
The constant $\mathcal{R}$ can now be absorbed in a redefinition of the potential 
\be \label{shify}
\Psi= -\frac{\dot a}{a} \epsilon_0\to \Psi= -\frac{\dot a}{a} \epsilon_0-\mathcal{R},
\ee
so that equation \eqref{ij} returns to its ($F=0$) Newtonian-gauge form\footnote{The Weinberg solution of the perturbation equations constructed here requires that the quantity $\mathcal{R}$ is a strict constant. However, the physical relevance of the existence of such a solution resides in the fact that it will have to approximate well suitable solutions for the Fourier modes of scalar perturbations in the super Hubble regime where space dependencies can be neclected.}. 
While all the other equations are in principle affected by the previous shift via their dependence on $\Psi$, a more precise analysis shows that his is not the case. On the one hand, equations \eqref{ident} 
%satisfied due to the Raychaudhuri equation of the background before demanding that $\partial_i R=0$ (this is obvious because 
and \eqref{lalila}  remain unchanged as they depend only on the time derivative of $\Psi$. On the other hand, 
equation \eqref{lapla} remains unchanged because the term $\nabla^2(\Psi-\dot{a} F)$---the only one that does not involve time derivatives of $\Psi$---initially vanishing is sent to $-\nabla^2 \mathcal{R}$ which is still vanishing as $\mathcal{R}$ is a constant.  Therefore, the shift of $\Psi$ by the constant $\mathcal{R}$, together with condition \eqref{marita}, does not change the form of the Newtonian gauge perturbation equations.
 
Note that, as mentioned above, the shift of $\Psi$ introduced in \eqref{uff} is not a gauge transformation.
Now we will see that the condition \eqref{marita} completely determines the physical solution of the Newtonian-gauge perturbation equations found via
the present procedure. In order to do that we write \eqref{marita}---using \eqref{gygy}---as a condition on $\epsilon_0$, namely
\be\label{uff}
 a \frac d {d\tau} \left(\frac {\epsilon_0} {a}\right)+2 \frac{\dot{a}} a\epsilon_0=-\mathcal{R}.
\ee
The solution of  \eqref{uff} is given by
\be\epsilon_0(\tau)=-\mathcal{R}\frac{ \int_{T}^\tau a(\tau') d\tau'}{a(\tau)},\ee
where an integration constant is hidden in the choice of initial time $T$.
Remarkably, by replacing $\epsilon_0$ in the expressions for  $\Psi$ and $\Phi$ we obtain for free their equality  imposed by equation \eqref{ij} for general modes in the absence of stresses ($\pi_S=0$), namely 
\be\label{huy1}
\boxed{\Phi=\Psi=-\mathcal{R}\left(1-H\frac{ \int_{T}^\tau a(\tau') d\tau'}{a(\tau)}\right).}
\ee
More precisely, we did not need to impose the physicality constraint $\Phi-\Psi=0$ imposed in \cite{Weinberg:2008zzc}. Here, it just follows from  
the consistency of the initial gauge transformation plus the shift in the definition of $\Psi$. 
For $\delta u$ one has
\be\label{huy2}
\boxed{\delta u_0=-\epsilon_0=\mathcal{R}\frac{ \int_{T}^\tau a(\tau') d\tau'}{a(\tau)},}
\ee
and for the matter perturbations we get 
\be\label{huy3}
\boxed{\delta\rho_{\alpha}=\dot \rho^0_{\alpha} \epsilon_0=-\dot \rho^0_{\alpha}\mathcal{R}\frac{ \int_{T}^\tau a(\tau') d\tau'}{a(\tau)}}
\ee
for any species $\alpha$. Indeed for any scalar quantity the solution would look the same.  One calls such type of perturbations adiabatic.

The adiabatic property follows from the fact that the change have been found via a special gauge transformation $\epsilon_0$ (in fact it can be interpreted as an infinitesimal time reparametrization for scalars and hence it affects all in the same universal way. This is of course a form of equivalence principle at play).
We modified the fields in two steps: first the previous gauge transformation, and second the shift by a constant of $\Psi$  (which restricts the time dependence of the gauge parameter $\epsilon_0(\tau)$).
Because the Newtonian gauge perturbation equations are invariant under the previous action, \eqref{huy1} and \eqref{huy2} define a  non-trivial\footnote{Very importantly, this transformation is not a gauge transformation because of the constant shift in $\Psi$. This is why gauge invariant observables will have non trivial values in this solution.} solution of the cosmological perturbation equations that is homogeneous. As such it must be a good approximation to solutions for modes with wavelengths much larger than $H^{-1}$. 

Finally, it is simple to check that a constant traceless tensor $D_{ij}$  is a zero mode solution for tensor modes \eqref{eqbox12}

\section{The difference with the standard paradigm where inhomogeneities arise from vacuum fluctuations.}\label{BD}

Here we discuss in more detail the difference of our model  with the more standard (by now textbook) accounts where the vacuum fluctuations in the quantum state of the 
inflaton are  the source of inhomogeneities. After all even when there is no inflaton field driving inflation, our model still has a scalar field degree of freedom
which  if  set (asymptotically in the far past) in the Bunch-Davies vacuum would have vacuum fluctuations analogous to that of the inflaton (indeed this is the idea in models of Higgs inflation). 

Some of the conceptual difficulties in interpreting such a paradigm has been discussed in \cite{Weinberg:2008zzc}. Here we will simply state that, in the absence of a theory of quantum gravity, the naturally available tool is that of semiclassical gravity where one replaces Eintein's equations by
\be\label{scg}
{\mathbf R}_{ab} - \frac{1}{2} {\mathbf R} g_{ab} = {8 \pi G} \braket{\psi |{\mathbf T}_{ab}|\psi},
\ee 
for some quantum state $\ket{\psi}$ of the matter living on a classical geometry. One sees immediately this approach would immediately lead (in the standard account) to no gravitational effects of vacuum fluctuations. More precisely, as cosmological perturbation theory is based on linearized gravity around the FLRW background $\braket{\psi |\delta{\mathbf T}_{ab}|\psi}=0$. For that reason one cannot interpret the standard account in terms of semiclassical gravity (which is not necessarily a problem) and needs to consider the simultaneous quantisation of matter and geometry in the framework of perturbation theory.  However,  difficulties arise when trying to account for the actual density fluctuations observed at the CMB from primordial vacuum fluctuations of matter fields and geometry because this requires, on the one hand, the interpretation of  quantum theory in the particularly thorny context of the universe as a whole with and additional assumption that quantum fluctuations turn into `real' classical fluctuations.  

The first key difference introduced by our model is that in our case fluctuations are generated in the state of the Higgs itself via the interaction of the (assumed) Planckian granularity and the scalar degrees of freedom. These fluctuations are semiclassical from the very beginning representing the imprint of the violation of the FLRW symmetries at the Planck scale. In our case the fluctuations are present in the semiclassical state of the Higgs $\ket{\psi}$ in the sense that $\braket{\psi |\delta{\phi}_{ab}|\psi}\not=0$ from the onset and consequently
\be
\braket{\psi |\delta{\mathbf T}_{ab}|\psi}\not=0.
\ee 
Thus, our model admits a semiclassical account corresponding to the linearized version of \eqref{scg}.
The state of the Higgs (via its interactions with the Planckian granularity) breaks the FLRW symmetry in contrast with the Bunch-Davies vacuum.
In our case, inhomogeneities are inherent in the Planckian substratum and simply transmitted to the scalar degree of freedom during inflation.
This completely eliminates the conceptual difficulties of the standard account and provides an alternative story which 
is consistent and has the appealing feature of linking inhomogeneities in the CMB with the Planckian fundamental discreteness 
predicted by several approaches to quantum gravity.
% COMPARACION DE FLUCTUACIONES
%One can compare the predictions of the two alternative perspectives when applied to our model when it comes to the form of the power spectrum of scalar perturbations:
%by following our proposal, or by, instead, assuming that the Higgs is assymptotically in the far past in the Bunch-Davies vacuum. In the second case (the standard approach) 
%the result for the amplitude of the power spectrum (here we follow Chapter 10.3 in \cite{Weinberg:2008zzc} which agrees with the standard treatment, for instance see \cite{Brandenberger:2003vk})  
%\be
%N^2_{\rm vac}=\frac{ 1} {4\pi^2 |\epsilon|} \frac{H_0^2}{m_p^2} \approx \frac{ 9} {64\pi^3} \frac{1}{\lambda^2 }
%\ee
%and the spectral index
%\be
%1-n_{\rm S}=2\delta+4\epsilon\approx \frac{8}{3} \lambda +\sO(\lambda^2)
%\ee
%where we have used the standard definitions of the slow roll parameters (evaluated in our model)
%\be
%\epsilon=-\frac{\dot H}{H^2}\approx \frac{16\pi }{9} \lambda^2  \ \ \ \ {\rm and } \ \ \ \  \delta=\frac{\ddot H}{2H \dot H}\approx \frac{4}{3}\lambda.
%\ee
%One easily sees that the previous results are incompatible with observations in the CMB. This puts clearly the important difference between the
%two different initial states for the scalar field.
%

{In our model the perturbations of the Higgs are born at horizon crossing and hence the state differs from the Bunch-Davies vacuum state: the `order parameter' revealing this difference is the expectation value $\braket{\psi |\delta{\phi}_{ab}|\psi}$ which vanishes in the Bunch-Davies vacuum but not in the present case. In the standard formulation the state of the inflaton is assumed to be given by the Bunch-Davies vacuum which is defined asymptotically in the far past introducing in this fashion the so-called trans-Planckian problem where initial conditions for the modes are given when their wavelengths are shorter than the Planck length. In our case the properties of the semiclassical state are defined at horizon crossing as discussed in Section \ref{sf}. Trans-Planckian modes play no role in our model.}

{ A clear way to sharply distinguish our state from the Bunch-Davis vacuum is the following: If, as in the standard paradigm, one would assume that evolution would be accurately dictated by the rules of quantum field theory on a curved classical spacetime (De Sitter spacetime) for trans-Planckian modes---a thought exercise which is physically inappropriate in our situation as one would be ignoring the quantum gravity effects which produce the semiclassical excitations in our state---the back-in-time evolution of our state would become singular in the asymptotic past. This is due to the fact that the semiclassical excitations produced by the granularity would be infinitely blue shifted by the De Sitter expansion towards the past and the state will deviate more and more from the (Bunch-Davies) vacuum. This is a simple way to illustrate the difference between the quantum state of the Higgs in our model from the choice made in the standard formulations. Of course, the singular behaviour to the past of our state is not a problem as it arises in the present exercise only when we ignore the mechanism of excitation and use the rules of quantum field theory beyond their regime of applicability.}

%See and include \cite{Burda:2015isa}.
%\subsection{On the possible fundamental origin of the fluctuations}
%
%ESTO NO ESTA REVISADO Y TIENE ECUACIONES QUE NO PUEDO JUSTIFICAR...
%\begin{figure}[h]
%%\begin{center}
%\centerline{\hspace{0.5cm} \(
%\begin{array}{c}
%\includegraphics[height=5cm]{feynman3.pdf} 
%\end{array} \ \ \ \ \  \ \ \ \ \  \ \ \ \ \  \ \ \ \ \  \ \ \ \ \  \ \ \ \ \  \ \ \ \ \  \ \ \ \ \   \begin{array}{c}
%\includegraphics[height=5cm]{feynman2.pdf} 
%\end{array}  \)} \caption{Higgs tree level interactions in the SM. The panel on the left represents the interaction of the Higgs with the $SU(2)$ gauge bosons allowing for the back reaction on the Higgs field. The panel on the right represents a generic fermion channel.} 
%\label{ll}
%%\end{center}
%\end{figure}
%
%\be
%\gamma\approx \exp\left(-\frac{4M_A}{M_\phi}\right)= \exp\left(-\frac{4 \sqrt{g} }{\sqrt{\lambda}}\right)
%\ee

\section{Speculations about some of the open questions}\label{openissues}

In this section we mention and discuss a few points that deserve further attention. We raise several questions 
here and propose possible tentative solutions. These open issues represent possible lines for future improvement 
of the ideas in this paper that we hope could be developed in the future.  

\subsubsection*{On the decay of the cosmological constant after the EW transition}

Among the few free dimensionless parameters entering our model there is $\beta$ which needs to be extremely small ($< 10^{-80}$) to produce a sufficiently long period of inflation:  the free parameter $\beta$ can be thought of as the free choice of the number of e-folds that take place in the scenario proposed during the inflationary period.  Thus the question of why $\beta$ is small is equivalent to the requirement encoded in equations like \eqref{postilla}. Note that such conditions are inequalities (inflation must be sufficiently long), and  thus $\beta$ does not enter any of the quantitative predictions of our proposal when considering observable imprints at the CMB.  Phrasing this in terms of $\beta$ note that its smallness is not, by itself, necessarily problematic in an effective description of a phenomenon that is emergent from the collective behaviour of tiny microscopic building blocks whose precise physics is not taken into account. Lacking such a fundamental description, one can try to find some possible guidance in dimensional analysis.  For instance recall that there was still a rescaling ambiguity in the definition of the unimodular time that actually 
rules the cosmological constant relaxation. Such possible rescaling  of the time variable $t_p$---introduced in \eqref{titita}---was encoded in the choice of a length scale $\ell_{\rm U}\gg \ell_p$.  Notice that this is quite reasonable as, in addition to Planck scale, there is another natural scale in the application of the cosmological principle to the region of interest of the universe: an IR scale $\ell_{\rm U}$ representing the extent of the `patch' of the universe that is well approximated by the ansatz geometry \eqref{metric} with homogeneous and isotropic background fields living on it. 

In terms of the time variable $t_p$ defined by \eqref{titita} with such IR scale, the relaxation is controlled by the `bare' value $\beta_0$ given by 
\be\label{petitese}
\beta=\beta_0\left(\frac{\ell_p}{\ell_{\rm U}}\right)^3.
\ee
Such reparametrization does not resolve the fine tuning problem and only shifts the issue of the smallness of $\beta$ into that of the largeness of $\ell_{\rm U}$.
However, it offers a new perspective pointing at the possibility of a physical mechanism where the size of the FLRW patch $\ell_U$  would stabilize the cosmological constant in essence by reducing diffusion. Such perspective suggests a long range quantum coherence mechanism (like for the collective behaviour in a Bose-Einstein condensates in relation to superfluidity) and offers a prospect for future analysis.

If such would be the role of $\ell_U$ this would also help resolving another question that necessarily arises when considering the instability of the cosmological constant possibly at play at the present (later stage) of the universe. That is: why is it that the present cosmological constant has not decayed yet by a similar relaxation? One advertised feature of our model  is that it proposes a new and different view on the question of why the cosmological constant can start at about its natural value and become basically zero quickly after the end of inflation. The remaining issue is how can it grow back to the value that is compatible with present observations. In \cite{Perez:2017krv, Perez:2018wlo} a model was proposed  (motivated by the same theoretical ideas as in this work) where the cosmological constant of the correct order of magnitude is generated due to diffusion during the electroweak transition. If the two proposals are to be consistent with each other then one would need a mechanism granting that the relaxation of the newly generated cosmological constant does not completely annihilates it by the present time. We notice that the phenomenological proposal \eqref{petitese}, characterizing the long scale coherence,  could possibly reconcile the two from the fact that the new IR (stabilizing) scale  $\ell^{\rm ew}_{\rm U}=a_{\rm ew} \ell_{\rm U}$ has expanded by the time of the electroweak transition. 
Therefore, even if the same relaxation mechanism would be at play after the EW transition, it could be sufficiently slow for the cosmological constant to persist basically unchanged  until present if $\ell_{\rm U}$ is sufficiently large to begin with. This follows from the fact that the change in unimodular time $\Delta t$ from the EW transition to today goes like 
$\Delta t \approx H_{\rm today}^{-1} a_{\rm today}^3$
(recall $a_0=1$ at the Planck initial time). Hence, a cosmological constant created at the electroweak time will last until today if
\be\label{conflict}\beta_0  \left(\frac{\ell_p} {\ell_{\rm U}}\right)^3\left(\frac{a_{\rm today}} {a_{\rm ew}}\right)^3 \frac{m_p}{H_{\rm today}}<1,
\ee 
i.e., we would be in a new inflationary regime for the new relaxing $\Lambda$.  Taking $\beta_0\approx 1$, the previous condition would require the initial coherence IR distance to be $\ell_U\ge10^{35}\ell_p= 1{\rm m}$. This appears as a huge initial region for our original bubble inflating to the present universe; at the same time we know and it has been often emphasised on various grounds that our universe requires extremely special initial conditions to accommodate its most basic features \cite{Penrose:1994de}. 

Even when the previous is perhaps the simplest speculation available with the details that our model offer, there could be other reasons for the relaxation process to change after the electroweak scale, rooted in some unknown quantum gravity mechanism that is no longer operational at such low energies. Such physics could be related to the role of the Higgs scalar in the whole picture. We notice that when the cosmological constant has relaxed to zero during the inflationary epoch, the Higgs scalar settles to its $V(\phi)=0$ configuration which certainly changes the coupling of this field with four volume in the effective action (having in mind that the interaction between fundamental four volume elements and matter could be the root of the diffusion mechanism that is central in our scenario). Other ideas explaining a possible `phase transition' that would make the relaxation stop after the electroweak scale are under investigation. 

\subsubsection*{The instability of the Higgs potential and quantum gravity}

In the present model the universe starts in a special state where the cosmological constant is of the order $m_p^2$ and the Higgs field is around the Hubble rate which itself is of the order of $m_p$. At such high values of $\phi_0$ the quartic coupling $\lambda$ is negative and---as we have seen in Section \ref{resultados}---this is exactly what is needed to explain the red tilt of the power spectrum of scalar perturbations.  However, this also implies that the Higgs field finds itself exactly in the instability region and is rolling towards higher values on the way to the Planck scale and beyond. 

Note however that the run-away behaviour is slow during the inflationary phase as the Hubble friction is very important due to the effect of a large cosmological constant (recall equation \eqref{47}).  When inflation ends the Hubble rate starts decaying and the instability becomes an issue. However, such conclusion only applies if one assumes that the standard model holds true beyond the Planck scale which is of course unreasonable. Deviations from the standard model should eventually become important as the Higgs field approaches $m_p$, and---even when it is hard to know what exactly that new physics would be in such regime (notice that even the standard QFT formulation on a curved background is expected to fail there)---it seems possible that such new physics could prevent the Higgs to roll to arbitrary high values. There are various models in the literature that try to render such conclusion more concrete (all sharing the limitation of the necessary reliable inputs from a quantum gravity theory). For instance, a non minimal coupling  of the Higgs with the geometry---which are necessary in models of Higgs inflation (see \cite{10.3389/fspas.2018.00050} for a review)---is shown to help stabilizing the Higgs up to about the Planck scale \cite{Espinosa:2015qea}. Other models predict stability at around the Planck scale \cite{Branchina:2014rva} by making assumptions on possible new physics. As an example,  a repulsive barrier at the Planck scale can arrise via $\phi^6$ and $\phi^8$ corrections of  the Higgs potential motivated by  grand-unified scenarios at $m_p$ \cite{Isidori:2001bm}. Such repulsive barrier at the Planck scale would only stop the Higgs scalar from rolling to arbitrary high transplankian scales. However, this by itself would not explain how the Higgs would eventually exit from that Planckian state and evolve towards the electro-weak minimum that produces the phenomenology of the standard model in accordance with the world we see around us. This problem resonates in some respects with the `gracefull exit' problem in models of Higgs inflation \cite{Bezrukov:2007ep, Bezrukov:2010jz}. Yet it is also different as, on the one hand,  in our model inflation is not driven by the Higgs, and, on the other hand, the diffusion of energy from the decaying cosmological constant raises the temperature of radiation back to close to the Planck temperature at the end of the inflationary era  (recall Figure \ref{fig:loglog} and the discussion in Section \ref{higgies},  equation \eqref{fine}). When temperature reaches Planckian values, at the onset of the radiation domination (recall Figure \ref{fig:loglog}),  the Higgs could thermalize decaying to the EW vacuum away from the instability scale (such possibility is explored in related scenarios in \cite{Espinosa:2015qea, Joti:2017fwe}). We are aware that a clear account of this is lacking in our scenario, for the usual reason that this part of the story concerns physics at the Planck scale.

\subsubsection*{On the validity of the semiclassical analysis}

The mechanism of generation  of structure in our model is based on the interaction of the Planckian granularity of quantum gravity with the low energy degrees of freedom encoded in the Higgs scalar field of the standard model. The analysis has been performed using the classical field equations for the scalar field evolving in a classical background. This is what one can do at the moment given the limitations of present quantum gravity theories to provide reliable calculation tools in such an extreme regime. The validity of semiclassical methods is an assumption of our analysis. Nevertheless, one must keep in mind that this limitation is shared by (and possibly more severe) in standard approaches where strong assumptions about trans--Planckian physics are customarily made. Note that in contrast there is no trans Planckian issue here. In our model, perturbations are born at the length scale $H^{-1}_0$ which is about the Planck length $\ell_p$. This is closer to the  the regime where the semiclassical treatment might become a reliable approximation.  

On a similar ground there is another issue that is common to various approaches and it is also shared by ours. This issue is sourced in the use of stochastic methods in conjunction with Einsteins equations and the difference between stochastic averages (satisfying some form of continuity equation compatible with the Bianchi identities or with the integrability conditions of unimodular gravity in our case) and the fact that individual realisations are not subjected in any clear fashion to such constraints. This implies that a single element of our stochastic ensemble does not follow the field equations of general relativity. This problem is often overlooked but it is present even in the standard paradigm of structure formation in inflation where quantum vacuum fluctuations are interpreted as classical stochastic fluctuations of an ensemble of realisations.   In our model the behaviour of the individual realisation that represents our universe follows a dynamics with should be describable via a more fundamental theory. Our mean field description is only effective and the possible conflict with the structure of Einstein's equations at the level of an individual element of the ensemble is to be resolved by quantum gravity.

\end{appendix}

%\bibliography{/Users/alejandroperez/Dropbox/REFS/referencias-master}
%\bibliographystyle{unsrt}
%\bibliographystyle{/Users/alejandroperez/Dropbox/REFS/bib-style}

\providecommand{\href}[2]{#2}\begingroup\raggedright\endgroup

\end{document}